\documentclass[12pt]{iopart}
\pdfoutput=1

\expandafter\let\csname equation*\endcsname\relax
\expandafter\let\csname endequation*\endcsname\relax
\usepackage{amsmath}

\usepackage{amssymb}
\usepackage{amsthm}

\usepackage{thmtools}
\usepackage{enumitem}

\usepackage[usenames,dvipsnames]{xcolor}
\usepackage{graphicx}
\usepackage{multirow}
\usepackage{float}
\usepackage{color}

\usepackage{xparse}

\usepackage[labelformat=simple]{subcaption}

\usepackage{todonotes}

\usepackage[hidelinks]{hyperref}
\hypersetup{
    colorlinks,
    linkcolor={black!100!black},
    citecolor={blue!100!black},
    urlcolor={blue!50!black}
}

\usepackage[normalem]{ulem}

\usepackage{pdflscape}

\usepackage{etoolbox}

\makeatletter
\def\@mkboth#1#2{}
\newlength\appendixwidth
\preto\appendix{\addtocontents{toc}{\protect\patchl@section}}
\newcommand{\patchl@section}{%
  \settowidth{\appendixwidth}{\textbf{Appendix }}%
  \addtolength{\appendixwidth}{1.5em}%
  \patchcmd{\l@section}{1.5em}{\appendixwidth}{}{\ddt}%
}
\makeatother

\makeatletter
\newrobustcmd{\fixappendix}{%
  \patchcmd{\l@section}{1.5em}{7em}{}{}%
  \patchcmd{\l@subsection}{2.3em}{7em}{}{}%
}
\makeatother

\newcommand \expe[2]	{\mathbb{E}_{#1}\left[#2\right]}
\newcommand \Mexpe[2]	{\mathbb{E}_{#1}^{(M)}\left[#2\right]}
\newcommand \gexpe[1]	{\expe{*}{#1}}
\newcommand \lexpe[1]	{\expe{rla}{#1}}
\newcommand \sexpe[1]	{\expe{rsa}{#1}}

\newcommand \aexpe[2]	{\hat{\mathbb{E}}_{#1}\left[#2\right]}

\newcommand \expet[2]	{\expe{#1}{\gamma_{#2}}}
\newcommand \Mexpet[2]	{\Mexpe{#1}{\gamma_{#2}}}
\newcommand \gexpet[1]	{\expet{*}{#1}}

\newcommand \sexpet[1]	{\expet{rsa}{#1}}
\newcommand \Msexpet[1]	{\Mexpet{rsa}{#1}}
\newcommand \expetw[1]	{\expet{#1}{\omega}}
\newcommand \Mexpetw[1]	{\Mexpet{#1}{\omega}}
\newcommand \gexpetw	{\expetw{*}}
\newcommand \lexpetw	{\expetw{rla}}
\newcommand \sexpetw	{\expetw{rsa}}
\newcommand \Msexpetw	{\Mexpetw{rsa}}
\newcommand \aexpet[2]	{\aexpe{#1}{\gamma_{#2}}}

\newcommand \aexpetw[1]	{\aexpet{#1}{\omega}}

\newcommand \asexpetw	{\aexpetw{rsa}}
\newcommand \var[2]		{\mathbb{V}_{#1}\left[#2\right]}
\newcommand \Mvar[2]	{\mathbb{V}_{#1}^{(M)}\left[#2\right]}
\newcommand \gvar[1]	{\var{*}{#1}}
\newcommand \lvar[1]	{\var{rla}{#1}}
\newcommand \svar[1]	{\var{rsa}{#1}}

\newcommand \Msvar[1]	{\Mvar{rsa}{#1}}
\newcommand \avar[2]	{\hat{\mathbb{V}}_{#1}\left[#2\right]}

\newcommand \asvar[1]	{\avar{rsa}{#1}}

\newcommand \probdelta[1]	{\delta_{#1}}
\newcommand \gprobdelta		{\probdelta{*}}
\newcommand \lprobdelta		{\probdelta{rla}}
\newcommand \sprobdelta		{\probdelta{rsa}}

\newcommand \probalphast[2]		{p_{#1,\:#2}}
\newcommand \gprobalphast[1]	{\probalphast{*}{#1}}

\newcommand \sprobalphast[1]	{\probalphast{rsa}{#1}}
\newcommand \probalphastw[1]	{\probalphast{#1}{\omega}}
\newcommand \gprobalphastw		{\probalphastw{*}}

\newcommand \sprobalphastw		{\probalphastw{rsa}}
\newcommand \aprobalphast[2]	{\hat{p}_{#1,\:#2}}

\newcommand \aprobalphastw[1]	{\aprobalphast{#1}{\omega}}

\newcommand \asprobalphastw		{\aprobalphastw{rsa}}

\newcommand \prob[1] {\mathbb{P}r\left[#1\right]}

\newcommand \bigO[1] {O\left(#1\right)}

\newcommand \oct {\text{oct}}

\def\arccot{\textrm{arccot}}

\newcommand \cross { \textbf{C} }

\NewDocumentCommand \cycle { O{n} }{ \mathcal{C}_{#1}}
\NewDocumentCommand \lintree { O{n} }{ \mathcal{L}_{#1}}
\NewDocumentCommand \complete { O{n} }{ \mathcal{K}_{#1}}
\NewDocumentCommand \compbip { O{n_1} O{n_2} }{ \mathcal{K}_{#1,#2}}

\newcommand{\sgn}{\mbox{sgn}}

\begin{document}

\title{Reappraising the distribution of the number of edge crossings of graphs on a sphere}
\author{Llu\'is Alemany-Puig$^{*,1}$, Merc\`e Mora$^2$ \& Ramon Ferrer-i-Cancho$^1$}
\address{
{\small $^1$Complexity \& Quantitative Linguistics Lab,} \\
{\small Departament de Ci\`encies de la Computaci\'o,} \\
{\small Laboratory for Relational Algorithmics, Complexity and Learning (LARCA),} \\ 
{\small Universitat Polit\`ecnica de Catalunya, }\\ 
{\small Barcelona, Catalonia, Spain.} \\
{\small $^2$ Group on Discrete, Combinatorial and Computational Geometry,}\\
{\small Departament de Matem\`atiques,}\\
{\small Universitat Polit\`ecnica de Catalunya, }\\
{\small Barcelona, Catalonia, Spain.}\\
{\small $^*$ Author to whom any correspondence should be addressed.}
}

\ead{lalemany@cs.upc.edu, merce.mora@upc.edu, rferrericancho@cs.upc.edu}

\begin{abstract}
Many real transportation and mobility networks have their vertices placed on the surface of the Earth. In such embeddings, the edges laid on that surface may cross. In his pioneering research, Moon analyzed the distribution of the number of crossings on complete graphs and complete bipartite graphs whose vertices are located uniformly at random on the surface of a sphere assuming that vertex placements are independent from each other. Here we revise his derivation of that variance in the light of recent theoretical developments on the variance of crossings and computer simulations. We show that Moon's formulae are inaccurate in predicting the true variance and provide exact formulae.
\end{abstract}



\maketitle

\tableofcontents

\section{Introduction}
\label{sec:intro}

The shape of our planet can be approximated by a sphere with a radius of about $3.9\cdot10^3$ miles. Many real transportation and mobility networks have vertices located on the surface of that sphere. These are examples of spatial networks, networks whose vertices are embedded in a space \cite{Barthelemy2018a}. In many transportation and mobility networks, the surface of the sphere is simplified as a projection on a plane \cite{Barthelemy2018a} while in some other cases, e.g., air transportation networks \cite{Li2004a,Guimera2004a,Guimera2005a}, such an approximation is often not possible due to the long distances involved.

When vertices are embedded in some space, edges may cross. While crossings are exceptional in many spatial networks to the point of being neglectable \cite{Barthelemy2018a}, crossings can also be scarce but not neglectable in one-dimensional layouts of certain networks: syntactic dependency and RNA secondary structures, where vertices are arranged linearly (distributed along a line) \cite{Ferrer2017a,Chen2009a}. The former are networks whose vertices represent words of a sentence and the edges represent syntactic dependencies between them. These have become the {\em de facto} standard to represent the syntactic structure of sentences in computational linguistics \cite{kubler09book} and the fuel of many quantitative studies \cite{Liu2017a, Temperley2018a}. In RNA secondary structures, vertices are nucleotides A, G, U, and C, and edges are Watson-Crick (A-U, G-C) and (U-G) base pairs \cite{Chen2009a}. In these one-dimensional networks, two edges cross whenever the endpoints' positions are interleaved in the sequence.

Statistical properties of $\cross$, the number of edge crossings of a graph $G$, have been studied in generic embeddings, denoted as $*$, that meet three mathematical conditions \cite{Alemany2018a}: (1) only independent edges can cross (edges that do not share vertices), (2) two independent edges can cross in at most one point, and (3) if several edges of the graph, say $e$ edges, cross at exactly the same point then the amount of crossings equals ${e \choose 2}=e(e-1)/2$. In our view, generic embeddings are two-fold: a space and a statistical distribution of the vertices in such space. In \cite{Moon1965a}, the space is the surface of a sphere while the distribution of the vertices on that surface is uniformly random. Compact formulae for the expectation and the variance have been obtained \cite{Alemany2018a}. Here we apply such a framework to revise the problem of calculating the distribution of $\cross$ in arrangements of vertices on the surface of a sphere. We use $\gexpe{\cross}=\gexpe{\cross(G)}$ to denote the expectation of the number of crossings $\cross$, and $\gvar{\cross}=\gvar{\cross(G)}$ to denote the variance of $\cross$ in a generic layout $*$.

In his pioneering research \cite{Moon1965a}, J. W. Moon analyzed the properties of the distribution of $\cross$ in uniformly random spherical arrangements (\textit{rsa}), where vertices are arranged on the surface of a sphere uniformly at random and independently from each other, and edges become geodesics on the sphere's surface. Specifically, Moon studied $\sexpe{\cross}$ and $\svar{\cross}$, the expectation and variance of $\cross$ in the random spherical layout, for two kinds of graphs: complete graphs of $n$ vertices, $\complete$, and complete bipartite graphs, $\compbip$, with $n_1$ vertices in one partition and $n_2$ vertices in the other. His derivations of $\sexpe{\cross}$ are straightforward. Borrowing the notation in \cite{Alemany2018a}, Moon obtained that the expectation of $\cross$ is
\begin{eqnarray*}
\sexpe{\cross} = q\sprobdelta,
\end{eqnarray*} 
where $q$ is the number of pairs of independent edges \cite{Piazza1991a} and $\sprobdelta$ is the probability that two independent edges cross. Indeed, $q$ is a handle for the size of the set $Q$, consisting of the pairs of independent edges of a graph $G$ \cite{Piazza1991a,Alemany2018a}. Thanks to 
\begin{eqnarray}
\label{eq:sphere:prob_edge_crossings}
\probdelta{rsa} = 1/8,
\end{eqnarray}
\begin{eqnarray*}
|Q(\complete)| = \frac{1}{2}{n \choose 2}{n - 2 \choose 2}
\end{eqnarray*}
and
\begin{eqnarray*}
|Q(\compbip)| = 2 {n_1 \choose 2}{n_2 \choose 2},
\end{eqnarray*}
Moon obtained
\begin{eqnarray*}
\sexpe{\cross(\complete)} &= \frac{1}{16} {n \choose 2}{n-2 \choose 2}, \\
\sexpe{\cross(\compbip)} &= \frac{1}{4}{n_1 \choose 2}{n_2 \choose 2}.
\end{eqnarray*}

Moon also derived formulae for the variance of $\cross$ for these two kinds of graphs, i.e.
\begin{equation*}
\Msvar{\cross(\complete)}
	=\frac{1}{2}{n \choose 2}{n - 2 \choose 2}
	\left[
		1 \cdot \frac{7}{64} +
		2{n-4 \choose 2} \cdot \frac{\pi^2 - 8}{64 \pi^2} +
		4(n-4) \cdot \frac{\pi^2 - 8}{64 \pi^2} +
		2 \cdot \frac{-1}{64}
	\right],
\end{equation*}
which simplifies to
\begin{eqnarray}
\label{eq:var_C:rsa:complete:Moons_raw}
\Msvar{\cross(\complete)}
	&=
	3{n \choose 4}
	\left[
		\frac{5}{64} + \frac{\pi^2 - 8}{64\pi^2}(n - 4)(n - 1)
	 \right]
\end{eqnarray}
and also
\begin{equation}
\label{eq:var_C:rsa:complete_bipartite:Moons_raw}
\Msvar{\cross(\compbip)}
	=
	\frac{1}{16 \pi^2} {n_1 \choose 2}{n_2 \choose 2}
	[(n_1 - 1)(n_2 - 1)(\pi^2 - 8) + 2(\pi^2 + 4)],
\end{equation}
where the superscript $(M)$ is used to distinguish Moon's work from our own derivations. Here we revise equations \ref{eq:var_C:rsa:complete:Moons_raw} and \ref{eq:var_C:rsa:complete_bipartite:Moons_raw} in light of computer simulations and a recently introduced theoretical framework to investigate $\gvar{\cross}$ \cite{Alemany2018a}.

This article is organized as follows. Section \ref{sec:var_C:gen_layout} summarizes the general mathematical framework for the calculation of $\gvar{\cross}$ \cite{Alemany2018a}, and section \ref{sec:var_C:rsa_ours} adapts it to the case of random spherical arrangements. In section \ref{sec:revision_Moon}, we review Moon's calculations, and compare them with our own with the help of numerical estimates of $\svar{\cross}$ in complete and complete bipartite graphs in sections \ref{sec:revision_Moon:complete} and \ref{sec:revision_Moon:complete_bipartite} respectively. These numerical estimates confirm the correctness of our derivations and show that Moon's (\ref{eq:var_C:rsa:complete:Moons_raw})-(\ref{eq:var_C:rsa:complete_bipartite:Moons_raw}) are inaccurate. Section \ref{sec:discussion} discusses our findings and attempts to shed light on the origins of the inaccuracy of $\Msvar{\cross(\complete)}$ and $\Msvar{\cross(\compbip)}$. Section \ref{sec:methods} details all the numerical methods involved in the numerical calculation of $\svar{\cross}$. This section is placed after the discussion to make the presentation of the main arguments more streamlined. 

\section{The variance of $\cross$ in generic layouts}
\label{sec:var_C:gen_layout}

In \cite{Alemany2018a}, $\cross$ was defined as a summation of pairwise crossings between independent edges, i.e.
\begin{eqnarray}
\label{eq:C}
\cross = \sum_{\{e_1,e_2\}\in Q} \alpha(e_1,e_2),
\end{eqnarray}
where $\alpha(e_1,e_2)$ is an indicator random variable that equals 1 whenever the independent edges $e_1$ and $e_2$ cross in the given layout. This definition was used to derive the expectation of $\cross$ as
\begin{eqnarray}
\label{eq:exp_C:general}
\gexpe{\cross} = q\gprobdelta,
\end{eqnarray}
where
\begin{eqnarray}
\label{eq:delta_prob_crossing}
\gprobdelta = \gexpe{\alpha(e_1,e_2)}
\end{eqnarray}
for two independent edges $e_1$ and $e_2$ embedded in the layout. Hereafter, an edge is a set of two vertices, denoted as $e=\{s,t\}=st$. Since $\alpha$ is an indicator random variable, $\gprobdelta$ is the probability that two independent edges cross in the given layout $*$. For example, in Moon's random spherical arrangement $\sprobdelta=1/8$ \cite{Moon1965a}. Therefore, using (\ref{eq:exp_C:general}) we obtain
\begin{eqnarray}
\label{eq:exp_C:rsa}
\sexpe{\cross} = \frac{1}{8}q.
\end{eqnarray}
Moreover, in uniformly random linear arrangements ({\em rla}), where the vertices of a graph are placed along a linear sequence uniformly at random, $\lprobdelta=1/3$ \cite{Ferrer2013d}, and then
\begin{eqnarray*}
\lexpe{\cross} = \frac{1}{3}q.
\end{eqnarray*}

\begin{table}
	\caption{The classification of the types of products $\alpha(e_1,e_2)\alpha(e_3,e_4)$ abstracting from the order of the elements of the pair $(\{e_1,e_2\},\{e_3,e_4\})\in Q\times Q$. The type $\omega$ is the result of concatenating of $\tau$ and $\phi$, $|\upsilon|$ is the number of different vertices of the type, $\tau=|\{e_1,e_2\}\cap\{e_3,e_4\}|$ and $\phi=|(e_1\cup e_2)\cap(e_3\cup e_4)|$. The form of every type of product is illustrated using $s,t,...,y,z$ to indicate distinct vertices ($st$ indicates the edge formed by vertices $s$ and $t$; the same for $uv$, $wx$,...). Types fully acknowledged by Moon are marked in bold.}
	\label{table:types_of_products}
	\begin{indented}
	\item[]
	\begin{tabular}{lllllllll}
		\br
		$\omega$	& $(\{e_1,e_2\},\{e_3,e_4\})$	& $|\upsilon|$	& $\tau$	& $\phi$\\
		\mr
		{\bf 00}	& $(\{st,uv\},\{wx,yz\})$		& 8				& 0			& 0		\\
		01			& $(\{st,uv\},\{sw,xy\})$		& 7				& 0			& 1		\\
		021			& $(\{st,uv\},\{su,wx\})$		& 6				& 0			& 2		\\
		022			& $(\{st,uv\},\{sw,ux\})$		& 6				& 0			& 2		\\
		03			& $(\{st,uv\},\{su,vw\})$		& 5				& 0			& 3		\\
		{\bf 04}	& $(\{st,uv\},\{su,tv\})$		& 4				& 0			& 4		\\ 
		{\bf 12}	& $(\{st,uv\},\{st,wx\})$		& 6				& 1			& 2		\\
		{\bf 13}	& $(\{st,uv\},\{st,uw\})$		& 5				& 1			& 3		\\
		{\bf 24}	& $(\{st,uv\},\{st,uv\})$		& 4				& 2			& 4		\\
		\br
	\end{tabular}
	\end{indented}
\end{table}

The same definition of $\cross$ was used again in \cite{Alemany2018a} to study the variance of $\cross$ in a general layout, similarly to the way Moon did for the particular case of random spherical arrangements \cite{Moon1965a}. In \cite{Alemany2018a}, the variance of $\cross$ was expressed compactly as a summation over products between graph-dependent terms, the $f_\omega$'s, and layout-dependent terms, the $\gexpetw$'s, i.e. 
\begin{eqnarray}
\label{eq:var_C:general}
\gvar{\cross} = \sum_{\omega\in\Omega} f_\omega \gexpetw.
\end{eqnarray}
Formally, the type of a product is obtained applying a function $\mathcal{T}$ on a pair $(\{e_1,e_2\},\{e_3,e_4\})\in Q\times Q$ and then \cite{Alemany2018a},
\begin{eqnarray*}
f_\omega =
	\sum_{q_1\in Q}
	\sum_{\substack{q_2\in Q \\ \mathcal{T}(q_1,q_2)=\omega}}
	1.
\end{eqnarray*}
The crux to understand (\ref{eq:var_C:general}) are the layout-dependent terms, $\gexpetw$, actually a shorthand for
\begin{eqnarray}
\label{eq:var_C:exp_gamma_omegas}
\gexpetw = \gexpe{\alpha(e_1,e_2)\alpha(e_3,e_4) \;|\; \mathcal{T}(\{e_1,e_2\},\{e_3,e_4\})=\omega} - \gprobdelta^2,
\end{eqnarray}
where $\omega\in\Omega$ is the type of the product $\alpha(e_1,e_2)\alpha(e_3,e_4)$ for $(\{e_1,e_2\},\{e_3,e_4\})\in Q\times Q$. The type of product is determined by the vertices forming the edges of $\{e_1,e_2\},\{e_3,e_4\}$ as explained in detail in table \ref{table:types_of_products}. The set of all distinct types of products is
\begin{eqnarray}
\label{eq:alpha_products:code_types}
\Omega = \{00,01,021,022,03,04,12,13,24\}
\end{eqnarray}
following the encoding of each type in table \ref{table:types_of_products}.

\begin{table}
	\caption{$a_\omega$ and $F_\omega$ as function of type $\omega$. $\lintree$ and $\cycle$ denote linear trees (or path graphs) and cycle graphs of $n$ vertices, respectively, and the operator $\oplus$ indicates the disjoint union of graphs.}
	\label{table:types_as_subgraph_counting}
	\begin{indented}
	\item[]
	\begin{tabular}{@{}llll}
		\br
		$\omega$ & $a_\omega$ & $F_\omega$ \\
		\mr
		$00$	&	$6$ & $\lintree[2]\oplus\lintree[2]\oplus\lintree[2]\oplus\lintree[2]$ \\
		$01$	&	$4$ & $\lintree[3]\oplus\lintree[2]\oplus\lintree[2]$ \\
		$021$	&	$2$ & $\lintree[4]\oplus\lintree[2]$ \\
		$022$	&	$4$ & $\lintree[3]\oplus\lintree[3]$ \\
		$03$	&	$2$ & $\lintree[5]$ \\
		$04$	&	$2$ & $\cycle[4]$ \\
		$12$	&	$6$ & $\lintree[2]\oplus\lintree[2]\oplus\lintree[2]$ \\
		$13$	&	$2$ & $\lintree[3]\oplus\lintree[2]$ \\
		$24$	&	$1$ & $\lintree[2]\oplus\lintree[2]$ \\
		\br
	\end{tabular}
	\end{indented}
\end{table}

\begin{figure}
	\centering
	\includegraphics{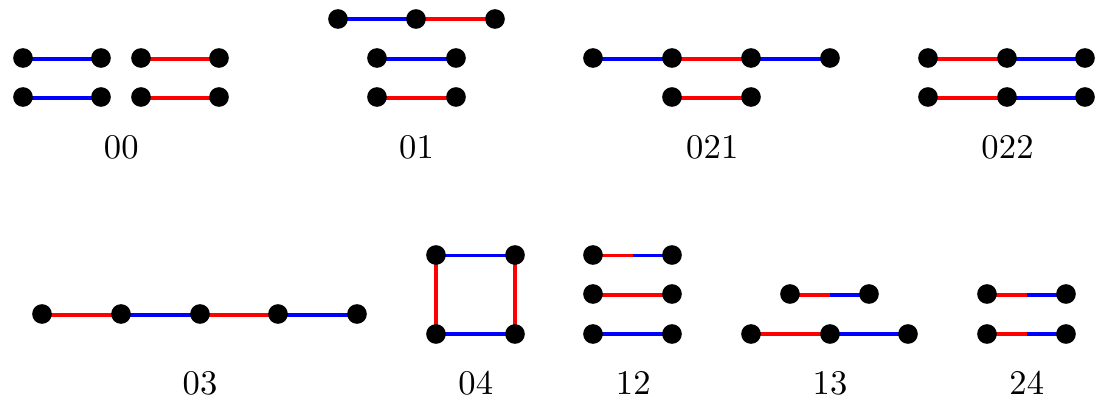}
	\caption{The subgraphs corresponding to each type of product of the form $\alpha(e_1,e_2) \alpha(e_3,e_4)$. The type $\omega\in\Omega$ (\ref{eq:alpha_products:code_types}) is indicated below them. In the product $\alpha(e_1,e_2) \alpha(e_3,e_4)$, $e_1$ and $e_2$ share the same color and $e_3$ and $e_4$ share another color. Equally-colored edges are to cross when calculating $\alpha(e_1,e_2) \alpha(e_3,e_4)$ and, by definition, belong to the same element of $Q$. Bi-colored edges (as in types $12$, $13$ and $24$) belong to both elements of $Q$.}
	\label{fig:types_subgraph}
\end{figure}

The amount of products $\alpha(e_1,e_2)\alpha(e_3,e_4)$ of type $\omega$ in the given graph $G$ satisfies \cite{Alemany2018a}
\begin{eqnarray*}
f_\omega = a_\omega n_G(F_\omega),
\end{eqnarray*}
where $a_\omega$ is a positive integer constant that depends only on $\omega$, and $n_G(F_\omega)$ is the number of subgraphs of $G$ isomorphic to the graph $F_\omega$ defined by the edges involved in the product $\alpha(e_1,e_2)\alpha(e_3,e_4)$ of type $\omega$. Figure \ref{fig:types_subgraph} depicts all these graphs. Table \ref{table:types_as_subgraph_counting} shows the values of the $a_\omega$ and the formal definition of each $F_\omega$.

For the sake of brevity, we use the shorthand
\begin{eqnarray}
\label{eq:var_C:prob_product_alphas}
\gprobalphastw = \gexpe{\alpha(e_1,e_2)\alpha(e_3,e_4) \;|\; \mathcal{T}(\{e_1,e_2\},\{e_3,e_4\})=\omega}.
\end{eqnarray}
Since $\alpha$ is an indicator random variable, $\gprobalphastw$ is the probability that both pairs of independent edges $\{e_1,e_2\},\{e_3,e_4\}\in Q$ cross in a generic embedding *.

Therefore, as it was concluded in \cite{Alemany2018a}, in order to calculate the variance of $\cross$ of a graph $G$ in a given layout $*$ one only needs to know the values of the $f_\omega$'s in $G$ (which are independent of the layout) and the values of the $\gexpetw$ in the given layout (which are independent of the graph).
The values of the $f_\omega$'s in complete and complete bipartite graphs, shown in table \ref{table:freqs:K__and__K_n1n2}, have allowed to derive expressions for the variance of $\cross$ that are valid for a generic embedding * \cite{Alemany2018a}. The substitution of the these values in (\ref{eq:var_C:general}) yields
\begin{eqnarray}
\label{eq:var_C:general:complete}
\gvar{\cross(\complete)}
	&= 3{n \choose 4} ((n - 4)(n - 5)(\gexpet{12} + 4(\gexpet{021} + \gexpet{022})) \nonumber\\
	&\phantom{=}	+ 4(n - 4)(\gexpet{13} + 2\gexpet{03}) \nonumber\\
	&\phantom{=}	+ 2\gexpet{04} + \gexpet{24})
\end{eqnarray}
and, likewise, 
\begin{eqnarray}
\label{eq:var_C:general:complete_bipartite}
\gvar{\cross(\compbip)}
&= 	 2(\gexpet{24} + \gexpet{04}) {n_1 \choose 2}{n_2 \choose 2} \nonumber\\
&\phantom{=}  + 12(\gexpet{03} + \gexpet{13})\left[ {n_1 \choose 3}{n_2 \choose 2} + {n_1 \choose 2}{n_2 \choose 3} \right] \nonumber\\
&\phantom{=} + 36(\gexpet{12} + \gexpet{022} + 2\gexpet{021}) {n_1 \choose 3}{n_2 \choose 3} \nonumber\\
&\phantom{=}  + 24\gexpet{022}\left[{n_1 \choose 4}{n_2 \choose 2} + {n_1 \choose 2}{n_2 \choose 4} \right].
\end{eqnarray}
A step further consists of instantiating the equations above replacing * by a uniformly random linear arrangements ({\em rla}). After calculating the values of the $\lexpetw$ and substituting them into (\ref{eq:var_C:general:complete}), one obtains \cite{Alemany2018a}   
\begin{eqnarray*}
\lvar{\cross(\complete)} = 0
\end{eqnarray*}
as expected. Interestingly, the same approach on (\ref{eq:var_C:general:complete_bipartite}) produces
\begin{eqnarray*}
\lvar{\cross(\compbip)} = \frac{1}{90}{n_1 \choose 2}{n_2 \choose 2}((n_1 + n_2)^2 + n_1 + n_2).
\end{eqnarray*}
Next section shows how equivalent results can be obtained when replacing $*$ by a random spherical arrangement, which turns out to be a more complex case.
 
\begin{table}
	\caption{Summary of the values of the $f_\omega$'s in complete graphs, $\complete$, and in complete bipartite graphs, $\compbip$. Values extracted from \cite[Table 5]{Alemany2018a}.}
	\label{table:freqs:K__and__K_n1n2}
	\begin{indented}
	\item[]
	\begin{tabular}{lll}
	\br
	$\omega$ & $f_\omega(\complete)$ & $f_\omega(\compbip)$ \\
	\mr
	00  & $630{n \choose 8}$ & $144{n_1 \choose 4}{n_2 \choose 4}$ \\
	01  & $1260{n \choose 7}$  & $144{n_1 \choose 4}{n_2 \choose 3} + 144{n_1 \choose 3}{n_2 \choose 4}$ \\
	021 & $360{n \choose 6}$ & $72{n_1 \choose 3}{n_2 \choose 3}$ \\
	022 & $360{n \choose 6}$ & $24{n_1 \choose 2}{n_2 \choose 4} + 24{n_1 \choose 4}{n_2 \choose 2} + 36{n_1 \choose 3}{n_2 \choose 3}$ \\
	03  & $120{n \choose 5}$ & $12{n_1 \choose 3}{n_2 \choose 2} + 12{n_1 \choose 2}{n_2 \choose 3}$ \\
	04  & $6{n \choose 4}$  & $2{n_1 \choose 2}{n_2 \choose 2}$ \\
	12  & $90{n \choose 6}$ & $36{n_1 \choose 3}{n_2 \choose 3}$ \\
	13  & $60{n \choose 5}$ & $12{n_1 \choose 3}{n_2 \choose 2} + 12{n_1 \choose 2}{n_2 \choose 3}$ \\
	24  & $3{n \choose 4}$ & $2{n_1 \choose 2}{n_2 \choose 2}$ \\
	\br
	\end{tabular}
	\end{indented}
\end{table}

\section{The variance of $\cross$ in spherical random arrangements}
\label{sec:var_C:rsa_ours}

Here we aim to calculate the values $\sexpetw$ so as to establish the foundations to derive an arithmetic expression for the variance of $\cross$ in uniformly random spherical arrangements of complete and complete bipartite graphs. Recall that, in this layout, vertices are distributed on the surface of a sphere uniformly at random, and edges become geodesics on that surface, i.e., great arcs (see section \ref{sec:methods:arc_arc_test} for a detailed account of what we consider valid arc-arc intersections). We use the $\sexpetw$'s to instantiate equations \ref{eq:var_C:general:complete} and \ref{eq:var_C:general:complete_bipartite} and then obtain arithmetic expressions for $\svar{\cross(\complete)}$ and $\svar{\cross(\compbip)}$ (section \ref{sec:var_C:rsa_ours:comp__comp_bip}). Each $\sexpetw$ is calculated via (\ref{eq:var_C:exp_gamma_omegas}): once $\sprobalphastw$ has been derived from (\ref{eq:var_C:prob_product_alphas}), $\sprobdelta^2$ (\ref{eq:delta_prob_crossing}) is subtracted.

We first calculate the $\sprobalphastw$ for the simplest cases. Thanks to \cite{Alemany2018a} we have that $\gprobalphast{24} = \gprobdelta$ and $\gprobalphast{00} = \gprobalphast{01} = \gprobdelta^2$. Since $\sprobdelta=1/8$ \cite{Moon1965a},
\begin{eqnarray*}
\sprobalphast{24} &= 1/8 \\
\sprobalphast{00} &= \sprobalphast{01} = 1/64
\end{eqnarray*}
and, following (\ref{eq:var_C:exp_gamma_omegas}), we obtain
\begin{eqnarray}
\sexpet{24} = 7/64				\label{eq:true_exp_gammas:24} \\
\sexpet{00} = \sexpet{01} = 0.	\label{eq:true_exp_gammas:00_01}
\end{eqnarray}
Furthermore \cite{Alemany2018a}, $\sprobalphast{04} = 0$ and then 
\begin{eqnarray}
\label{eq:true_exp_gammas:04}
\sexpet{04}=-1/64.
\end{eqnarray}
Notice that, given a pair of edges $\{st,uv\} \in Q$, we can form a type 04 combining $\{st,uv\}$ with $\{sv,tu\}$ or $\{su,tv\}$, which gives two configurations of type 04, $(\{st,uv\}, \{sv,tu\})$ and $(\{st,uv\}, \{su,tv\})$. For $\sprobalphast{04} > 0$, we need both indicator variables $\alpha$ to be $1$ for each pair of edges. However, if $\alpha(st,uv)=1$ then it is impossible that $\alpha(su,tv)=1$ or that $\alpha(sv,tu)=1$.

So far we have calculated $\sprobalphastw$ and $\sexpetw$ for $\omega\in\{00,01,04,24\}$ using arguments that can be applied to most layouts. Now we move on to $\sprobalphastw$ for $\omega\in\{021,022,03,12,13\}$ using an {\em ad hoc} approach for the spherical layout that is based on spherical trigonometry and integration over surfaces. Such surfaces are delimited by the edges that make up type $\omega\in\Omega$ (table \ref{table:types_of_products}). 

We proceed gradually towards such aim. First, we introduce the relevant background from spherical trigonometry (section \ref{sec:spherical_geometry}). Second, we propose a derivation for $\sprobdelta = 1/8$ that is more detailed than that of Moon \cite{Moon1965a} (section \ref{sec:probability_of_crossing}).
Finally, we proceed with the remainder of types of products namely $12$, $021$, $13$, $03$, $022$ (section \ref{sec:expectations}).

\subsection{Spherical trigonometry}
\label{sec:spherical_geometry}

We first recall some definitions and properties of spherical trigonometry \cite{Granville1908a}. Henceforth, we assume a unit-radius sphere. Let $A$, $B$ and $C$ be three points on a sphere of center $O$, such that $A$, $B$ and $C$ do not lie all together on a plane containing $O$.

\begin{figure}
	\centering
	\includegraphics[scale=0.5]{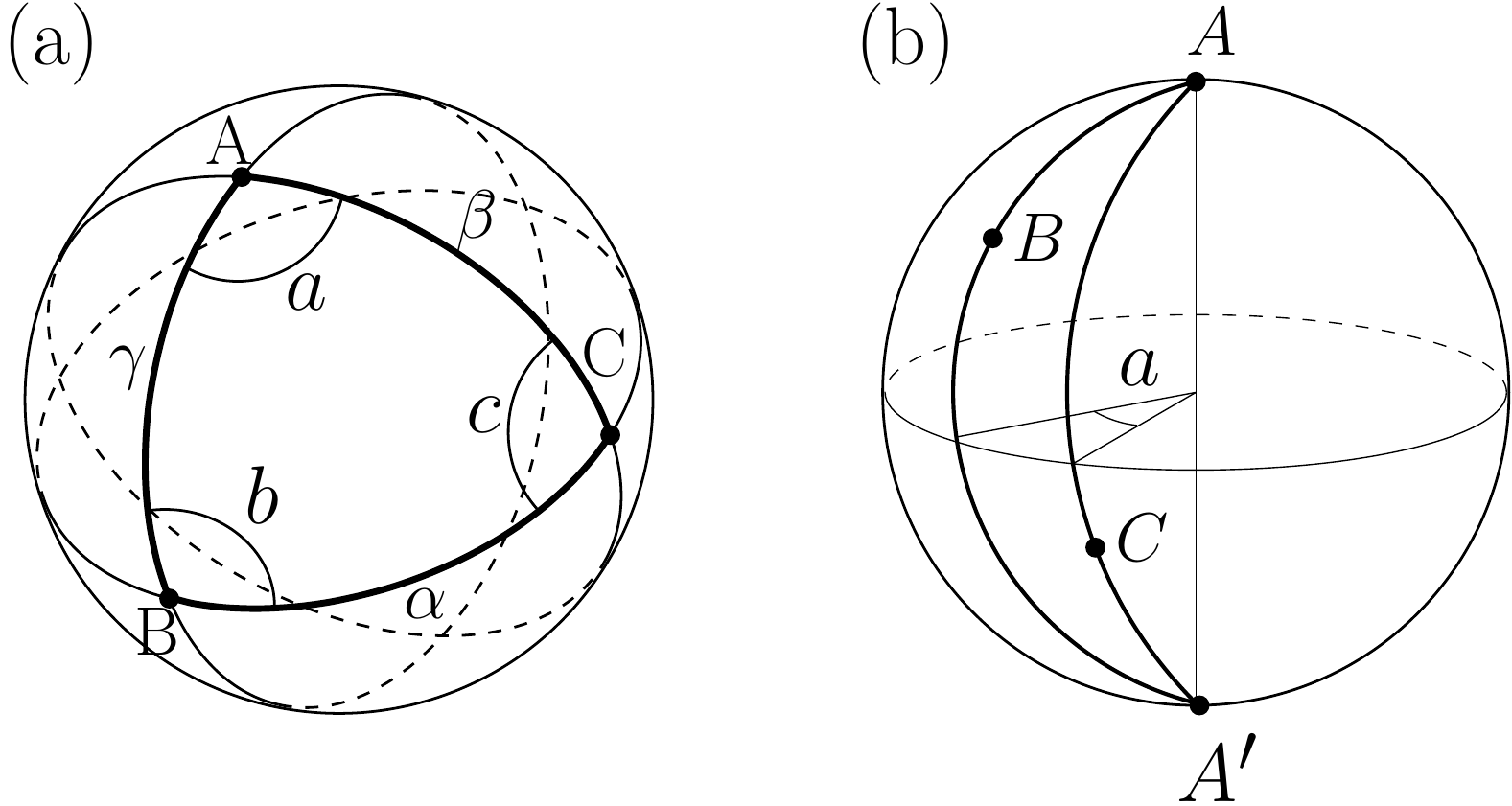}
	\caption{(a) A spherical triangle with vertices $A$, $B$ and $C$; sides $\alpha$, $\beta$ and $\gamma$, and angles $a$, $b$ and $c$. (b) The lune $w(A;B,C)$ of angle $a$.}
	\label{fig:wedge}
\end{figure}

Points $A$, $B$, $C$ define a \emph{spherical triangle}, denoted by $tr(A,B,C)$, whose vertices are $A$, $B$ and $C$, and whose sides are the geodesics joining $A$ with $B$, $B$ with $C$, and $C$ with $A$. Let $\alpha$, $\beta$ and $\gamma$ denote the length of the sides $BC$, $AC$ and $AB$, respectively (figure \ref{fig:wedge}(a)). 

For every point $P$, let $P'$ denote its antipodal vertex, that is, $P$ and $P'$ are diametrically opposite. Thus, the segment $\overline{PP'}$ goes through the center $O$. The semicircle on the sphere containing $A$, $A'$ and $B$, and the semicircle on the sphere containing $A$, $A'$ and $C$ delimit two disjoint regions on the sphere. 
The {\em lune} $w(A;B,C)$ is the one with smaller area.
The angle of $w(A;B,C)$ is the angle in $(0,\pi)$ defined by the planes containing the semicircles delimiting the lune (figure \ref{fig:wedge}(b)). The angles at vertices $A$, $B$ and $C$ of a spherical triangle $tr(A,B,C)$ are the angles $a$, $b$ and $c$ in $(0,\pi)$ of the lunes $w(A;B,C)$, $w(B;A,C)$ and $w(C;A,B)$, respectively (figure \ref{fig:wedge}(a)).

With this notation, the following formula relates the lengths $\alpha$, $\beta$ of two sides with the angles $b$ and $c$ of the spherical triangle $tr(A,B,C)$
\begin{eqnarray}
\label{eq:cotangentes}
\cot(\beta) \sin (\alpha)=\cos(\alpha) \cos(c)+\sin(c)\cot(b).
\end{eqnarray}
Using this equation, from the length of two sides and the angle at the shared vertex of a spherical triangle, the remaining angles can be calculated. Concretely,
\begin{eqnarray*}
\cot(b) = \frac{\cot(\beta) \sin (\alpha)-\cos(\alpha) \cos(c)}{\sin(c)}
\end{eqnarray*}
and, analogously,
\begin{eqnarray*}
\cot(a) =\frac{\cot(\alpha) \sin (\beta)-\cos(\beta) \cos(c)}{\sin(c)}.
\end{eqnarray*}
Since this relation is used often in our calculations, let us define a function $g$ such that for any real numbers $x,y,z\in (0,\pi)$,
\begin{eqnarray}
\label{eq:g_function}
g(x,y,z)=\arccot \left( \dfrac{\cot(x) \sin (y)-\cos(y) \cos(z)}{\sin (z)} \right).
\end{eqnarray}
%
Let $S(R)$ denote the area of a region $R$ of the sphere of radius 1.  
It is well-known that  the area of a lune is $S(w(A;B,C))=2a$, where $a$ is the angle of the lune, and the area  of a spherical triangle is $S(tr(A,B,C))=a+b+c-\pi$, where $a$, $b$ and $c$ are the angles at vertices $A$, $B$ and $C$, respectively.

\subsection{The probability that two edges cross}
\label{sec:probability_of_crossing}

\begin{figure}
	\centering
	\includegraphics[width=0.5\textwidth]{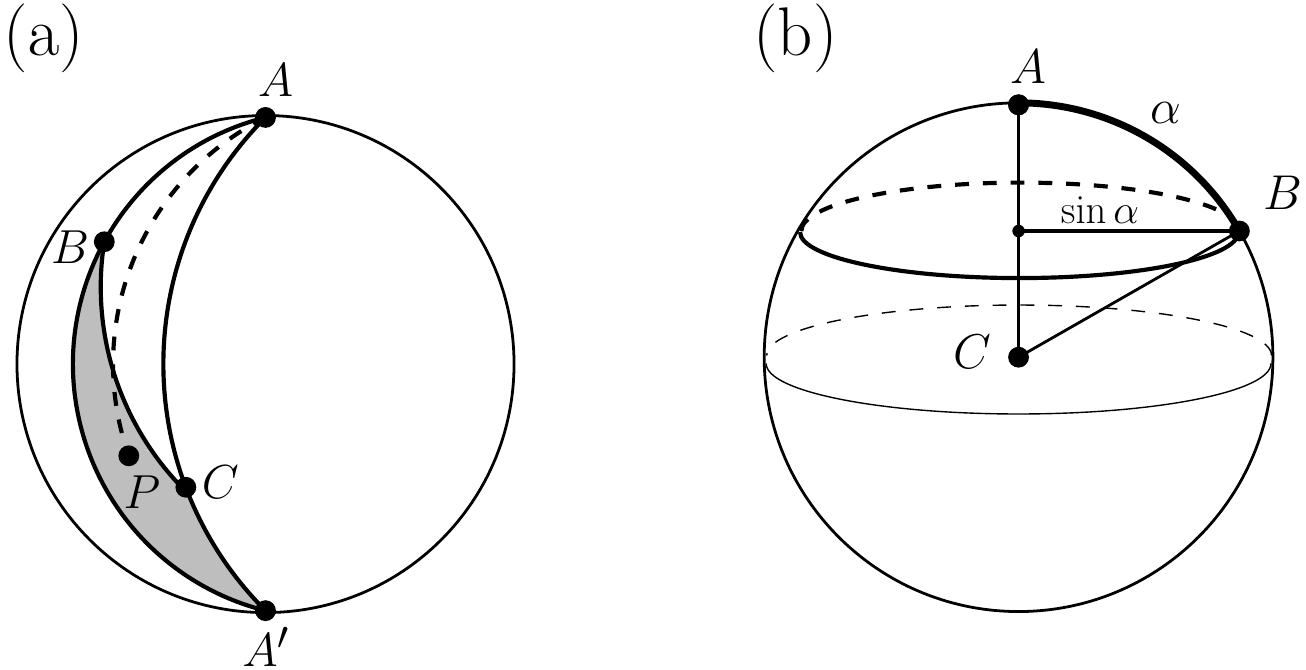}		
	\caption{(a) Given three points $A$, $B$ and $C$, and a random point $P$, the probability that the geodesic $AP$ crosses the side $BC$ is proportional to the area of the spherical triangle $tr(A',B,C)$. (b) Given two random points $A$ and $B$, the probability that the geodesic $AB$ has length $\alpha$ is proportional to the length of the circle that goes through $B$ and lying on the plane perpendicular to the radius $CA$; notice that this circle has radius $\sin \alpha$ if the radius of the sphere is 1.}
	\label{fig:probwedge_densitat}
\end{figure}

Let  $A$, $B$ and $C$ be three fixed points on the sphere and let $P$ be a random point. The geodesic $AP$ crosses  $BC$ if and only if $P$ lies in the spherical triangle $tr(A',B,C)$ (figure \ref{fig:probwedge_densitat}(a)). Hence, the probability of this occurring is the area of the spherical triangle $tr(A',B,C)$ divided by the area of the sphere's surface,
\begin{align}
\prob{\hbox{$AP$ and $BC$ cross}}
	&= \frac{S(tr(A',B,C))}{4\pi}=\frac{S(w(A;B,C))-S(tr(A,B,C))}{4\pi}\nonumber \\
	&= \frac{2a-(a+b+c-\pi)}{4\pi}=\frac{a-b-c+\pi}{4\pi}. \label{eq:probability_of_crossing}
\end{align}


Let $\alpha$ denote the length of the geodesic defined by two random points on the sphere. The density function of $\alpha $ is 
\begin{eqnarray}\label{eq:arclength}
f(\alpha)=\frac 12 \sin (\alpha). 
\end{eqnarray}
Indeed, we know that 
$\int_0^\pi f(\alpha)=1$ and that $f(\alpha)$ must be proportional to the length of the circle obtained by intersecting the plane containing one of the points and perpendicular to the line that goes through the other point and the center of the sphere (figure \ref{fig:probwedge_densitat}(b)). Taking into account these facts, we conclude that $f(\alpha)$ satisfies (\ref{eq:arclength}).

Let $s_1$ be a geodesic of length $\alpha$. The probability that two random points lie on the different hemispheres determined by $s_1$ is $1/2$, and the probability that the geodesic determined by two random points on different hemispheres cross a given arc of length $\alpha$ is $\alpha/2\pi$. Hence, the probability that a geodesic $s_2$ defined by two random points crosses a geodesic  $s_1$ of length $\alpha$ is
\begin{eqnarray}\label{eq:crossesgeodesicalpha}
 \prob{s_2\textrm{ crosses }s_1 | \textrm{ length of }s_1\textrm{ equal to }\alpha }= \alpha/4\pi .
\end{eqnarray}

Hence, the probability that two random edges on the sphere cross is, as already derived by Moon (\ref{eq:sphere:prob_edge_crossings}),
\begin{eqnarray}\label{eq:independentEdges}
\sprobdelta
	= \int^{\pi}_{0} \frac {\alpha}{4\pi}\, f(\alpha) d\alpha= \int^{\pi}_{0} \frac {\alpha}{8\pi}\, \sin (\alpha) d\alpha
	= 1/8.
\end{eqnarray}

\subsection{The $\sprobalphastw$'s and the $\sexpetw$'s}
\label{sec:expectations}



%
Let $(\{e_1,e_2\},\{e_3,e_4\})\in Q\times Q$ be an element  of type $\omega \in \Omega$ as described in table~\ref{table:types_of_products} (see also figure~\ref{fig:types_subgraph}).  Below we calculate $\sprobalphast{\omega}$ for every $\omega \in \{ 12, 021, 13, 03, 022 \}$.
Table \ref{table:types_probs_sphere} summarizes the results on $\sprobalphastw$ that are derived next analytically and confirms the accuracy of the results with the help of computer simulations. For a better understanding of the explanations below, we refer the reader to table \ref{table:types_of_products} (where we describe the types of products following \cite{Alemany2018a}), and figure \ref{fig:types_subgraph} (that illustrates the graphs characterizing each type).

\paragraph{Case $\omega={12}$.}
Let  $e=e_1=e_3$.  By (\ref{eq:arclength}) and (\ref{eq:crossesgeodesicalpha}), the probability that two random edges $e_2$ and $e_4$ cross the edge $e$ of length $\alpha$
is
\begin{eqnarray*}
\sprobalphast{12}
	= \int_0^{\pi} \left(\frac{\alpha}{4\pi}\right)^2 f(\alpha) \, d\alpha
	= \frac{1}{32\pi^2 } \int_0^{\pi} \alpha ^2 \sin (\alpha) \, d\alpha
	= \frac{\pi^2 - 4}{32\pi^2},
\end{eqnarray*}
and then
\begin{eqnarray}
\label{eq:true_exp_gammas:12}
\sexpet{12} = \frac{\pi^2 - 8}{64\pi^2}.
\end{eqnarray}

\begin{figure}
	\centering
	\includegraphics[width=0.8\textwidth]{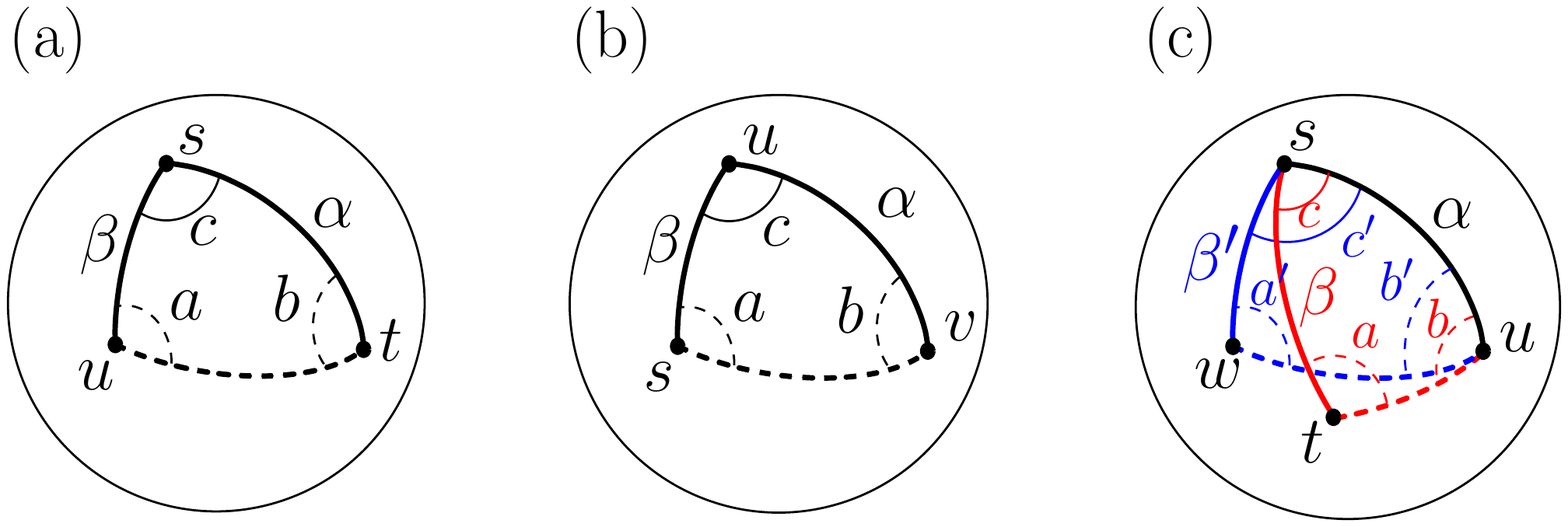}	
	\caption{(a) Types $021$  and $13$: the relative position of points $s$, $t$ and $u$ is determined by parameters $\alpha$, $\beta $ and $c$. (b) Type $03$: the relative position of points $s$, $u$ and $v$ is determined by parameters $\alpha$, $\beta $ and $c$. (c) Type $022$: the relative position of points $s$, $t$, $u$ and $w$ is determined by parameters $\alpha$, $\beta $, $\beta '$, $c$ and $c'$.  The spherical triangles $tr(s,u,w)$ and $tr(s,u,t)$ are drawn in black-blue and in black-red, respectively.}
	\label{fig:type03}
\end{figure}

\paragraph{Case $\omega={021}$.}
Recall that $e_1=st$, $e_2=uv$, $e_3=su$ and $e_4=wx$, with $s,t,u,v,w,x$ pairwise distinct. The relative position of $s$, $t$ and $u$ can be given by three independent parameters: the length $\alpha \in (0,\pi)$ of the geodesic $st$, the length $\beta\in (0,\pi)$ of the geodesic $su$ and the angle $c\in (-\pi, \pi)$ of the lune  $w(s;t,u)$ (figure \ref{fig:type03} (a)). 

On the one hand, given a random point $v$, 
the probability that the edge $e_2=uv$ crosses the edge $e_1=st$ whenever $c\in (0,\pi)$ can be derived using~(\ref{eq:probability_of_crossing}) for the triangle $tr(A,B,C)$ when $A=u$, $B=t$ and $C=s$.
Besides, this probability is the same for the opposite angle $-c$. 
On the other hand, by  (\ref{eq:crossesgeodesicalpha}), the probability that the edge $e_4$ crosses the edge $e_3$ of length $\beta$ is $\beta/(4\pi)$. Therefore,
\begin{align*}
\sprobalphast{021}
	&= 2 \iiint_0^{\pi} \frac{a-b-c+\pi}{4\pi} \cdot \frac \beta{4\pi} \, f(\alpha) f(\beta) \frac 1{2\pi}\, d\alpha \, d\beta \, dc \\
	&= 2 \iiint_0^{\pi} \frac{g(\alpha,\beta,c)-g(\beta,\alpha,c)-c+\pi}{4\pi} \cdot  \frac \beta{32 \pi^2}  \sin(\alpha) \sin (\beta) \, d\alpha \, d\beta \, dc \\
	&\approx
	0.013 
\end{align*}
and thus
\begin{eqnarray}
\label{eq:true_exp_gammas:021}
\sexpet{021} = \sprobalphast{021} - \frac{1}{64} \approx -0.003. 
\end{eqnarray}

\paragraph{Case $\omega={13}$.}
Recall that $e_1=e_3=st$, $e_2=uv$ and $e_4=uw$, with $s,t,u,v,w$ pairwise distinct. As in the preceding case, the relative position of $s$, $t$ and $u$ can be given by three independent parameters: the length $\alpha \in (0,\pi)$ of the geodesic $st$, the length $\beta\in (0,\pi)$ of the geodesic $su$ and the angle $c\in (-\pi, \pi)$ of the lune $w(s;t,u)$ (figure \ref{fig:type03} (a)).  
%
Moreover, given a random point $v$ (resp. $w$), 
the probability that the edge $e_2=uv$  (resp. $e_4=uw$) crosses the edge $e_1=st$ whenever $c\in (0,\pi)$ can be derived using~(\ref{eq:probability_of_crossing}) for the triangle $tr(A,B,C)$ when $A=u$, $B=t$ and $C=s$.
Besides, the probability of crossing is the same for the opposite angle $-c$.
Therefore, 
%
%
\begin{align*}
\sprobalphast{13}
	&= 2 \iiint_0^{\pi} \left( \frac{a-b-c+\pi}{4\pi} \right)^2 \, f(\alpha) f(\beta) \frac 1{2\pi} \, d\alpha \, d\beta \, dc \\
	&= 2 \iiint_0^{\pi} \left( \frac{g(\alpha,\beta,c)-g(\beta,\alpha,c)-c+\pi}{4\pi}\right)^2  \frac 1{8\pi} \sin(\alpha) \sin (\beta) \, d\alpha \, d\beta \, dc \\
	&\approx 0.031 
\end{align*}
and thus
\begin{eqnarray}
\label{eq:true_exp_gammas:13}
\sexpet{13} = \sprobalphast{13} - \frac{1}{64} \approx 0.016. 
\end{eqnarray}

\paragraph{Case $\omega = {03}$.}
Assume that $e_1=st$, $e_2=uv$, $e_3=su$ and $e_4=vw$, with $s,t,u,v,w$ pairwise distinct. Similarly as in the preceding case, the relative position of $u$, $v$ and $s$ can be given by three independent parameters: the length $\alpha $ of the geodesic $uv$, the length $\beta$ of the geodesic $us$ and the angle $c$ of the lune $w(u;v,s)$ (figure \ref{fig:type03}(b)). 
Moreover, given a random point $t$, the probability that the edge $e_1=st$ crosses $e_2=uv$ whenever $c\in (0,\pi)$ can be derived using~(\ref{eq:probability_of_crossing}) for the triangle $tr(A,B,C)$ when $A=s$, $B=v$ and $C=u$. Analogously, 
given a random point $w$, the probability that the edge $e_4=vw$ crosses $e_3=su$ can be derived using~(\ref{eq:probability_of_crossing}) for the triangle $tr(v,s,u)$.
Therefore, 
\begin{align*}
\sprobalphast{03}
	&= 2 \iiint_0^{\pi} \left( \frac{a-b-c+\pi}{4\pi} \right) \left( \frac{b-a-c+\pi}{4\pi} \right)\, f(\alpha) f(\beta) \, \frac 1{2\pi}  \, d\alpha \, d\beta \, dc \\
	&= 2 \iiint_0^{\pi}
	   \left( \frac{g(\alpha,\beta,c)-g(\beta,\alpha,c)-c+\pi}{4\pi} \right)
	   \left( \frac{g(\beta,\alpha,c)-g(\alpha,\beta,c)-c+\pi}{4\pi} \right) \\ 
	&\phantom{==}\cdot \frac{1}{8\pi} \sin(\alpha) \sin (\beta) \, d\alpha \, d\beta \\
	&\approx 0.01. 
\end{align*}
Thus 
\begin{eqnarray}
\label{eq:true_exp_gammas:03}
\sexpet{03} = \sprobalphast{03} - \frac{1}{64} \approx -0.0052. 
\end{eqnarray}
  

\paragraph{Case $\omega={022}$.}
Recall that $e_1=st$, $e_2=uv$, $e_3=sw$ and $e_4=ux$, with $s,t,u,v,w,x$ pairwise distinct. The relative position of points $s$, $t$, $u$ and $w$ can be given by 5 independent parameters, $\alpha$, $\beta$, $\beta'$,  $c$, and $c'$, where $\alpha $, $\beta$ and  $\beta'$
are the lengths of the geodesics $su$, $st$ and $sw$, respectively; $c$ is the angle of the lune $w(s;t,u)$; and $c'$ is the angle of the lune $w(s;u,w)$, with $c,c'\in (-\pi,\pi)$ (figure \ref{fig:type03}(c)).


Let $a$, $b$ denote the angles at vertices $t$ and $u$, respectively, of the spherical triangle $tr(t,u,s)$ and let $a'$, $b'$ denote the angles at vertices $w$ and $u$, respectively, of the spherical triangle $tr(w,u,s)$. For $c\in (0,\pi)$ and $c'\in (0,\pi)$, given two random points $v$ and $x$, the probability that $e_2=uv$ crosses $e_1=st$ and the probability that $e_4=ux$ crosses $e_3=sw$ can be calculated using (\ref{eq:probability_of_crossing}) for the triangles $tr(t,u,s)$ and  $tr(w,u,s)$, respectively. Since the probability of crossing is the same for the opposites angles of $c$ and $c'$, we derive that
\begin{align*}
\sprobalphast{022}
	&= 4 \idotsint_0^{\pi} \left( \frac{b-a-c+\pi}{4\pi} \right) \left( \frac{b'-a'-c'+\pi}{4\pi} \right) \\
	&\phantom{==}f(\alpha) f(\beta) f(\beta') \left( \frac 1{2\pi} \right)^2\, d\alpha \, d\beta \,  d\beta' \, dc \, dc' \\
	&= 4 \idotsint_0^{\pi} 
	   \left( \frac{g(\beta,\alpha,c)-g(\alpha,\beta,c)-c+\pi}{4\pi}\right)
	   \left( \frac{g(\beta',\alpha,c')-g(\alpha,\beta',c')-c'+\pi}{4\pi}\right) \\ 
	&\phantom{==}\frac 1{32 \pi^2} \sin (\alpha)\sin (\beta) \sin (\beta') \, d\alpha \, d\beta \, d\beta' \, dc \, dc' \\
	&\approx 0.019. 
\end{align*}
Thus
\begin{eqnarray}
\label{eq:true_exp_gammas:022}
\sexpet{022} = \sprobalphast{022} - \frac{1}{64} \approx 0.0029. 
\end{eqnarray}

The values of the integrals have been approximated as explained in section \ref{sec:methods}.


\begin{table}
	\caption{$\sprobalphastw$ is the theoretical probability that $\alpha(e_1,e_2)\alpha(e_3,e_4)=1$ for a product of type $\omega$ (\ref{eq:var_C:prob_product_alphas}). $\asprobalphastw$ is the corresponding numerical estimate via computer simulation (section \ref{sec:methods}). The last column points to the equation where the exact value of $\sprobalphastw$ is given. $\sprobalphastw$ is provided in two forms: exact value (types 00, 01, 04, 12 and 24) or approximate value by solving numerically the integrals in section \ref{sec:expectations} (types 021, 022, 03 and 13).}
	\label{table:types_probs_sphere}
	\begin{indented}
	\item[]
	\begin{tabular}{llll}
		\br
		$\omega$	& $\sprobalphastw$							& $\asprobalphastw$	& Equation \\
		\mr
		00			& 1/64 = $0.015625$							& 0.015625			& \ref{eq:true_exp_gammas:00_01} 
		\\
		01			& 1/64										& 0.015626			& \ref{eq:true_exp_gammas:00_01} 
		\\
		021			& $\approx$ 0.012665						& 0.012670			& \ref{eq:true_exp_gammas:021} \\
		022			& $\approx$ 0.018566						& 0.018581			& \ref{eq:true_exp_gammas:022} \\
		03			& $\approx$ 0.010401						& 0.010417			& \ref{eq:true_exp_gammas:03} \\
		04			& 0											& 0					& \ref{eq:true_exp_gammas:04} \\ 
		12			& $(\pi^2 - 4)/32\pi^2 \approx 0.018585$	& 0.018581			& \ref{eq:true_exp_gammas:12} \\
		13			& $\approx$ 0.031265						& 0.031251			& \ref{eq:true_exp_gammas:13} \\
		24			& 1/8										& 0.125001			& \ref{eq:true_exp_gammas:24} 
		\\
		\br
	\end{tabular}
	\end{indented}
\end{table}

\subsection{The variance of $\cross$ in complete and complete bipartite graphs}
\label{sec:var_C:rsa_ours:comp__comp_bip}

By substituting the values of the $\sexpetw$'s  (table \ref{table:types_probs_sphere}) into equations \ref{eq:var_C:general:complete} and \ref{eq:var_C:general:complete_bipartite}, we obtain, respectively,
\begin{eqnarray}
\label{eq:var_C:rsa:complete:ours}
\svar{\cross(\complete)}
	&\approx
	3{n \choose 4}
	\left[
		(n - 4)(0.0029(n - 5) + 0.021) + \frac{5}{64}
	\right]
\end{eqnarray}
and
\begin{eqnarray}
\label{eq:var_C:rsa:complete_bipartite:ours}
\svar{\cross(\compbip)}
&\approx
\frac{3}{16}{n_1 \choose 2}{n_2 \choose 2} - 0.00068{n_1 \choose 3}{n_2 \choose 3} \nonumber\\	
&\phantom{=}  + 0.12\left[ {n_1 \choose 3}{n_2 \choose 2} + {n_1 \choose 2}{n_2 \choose 3} \right] \nonumber\\
&\phantom{=}  + 0.07\left[{n_1 \choose 4}{n_2 \choose 2} + {n_1 \choose 2}{n_2 \choose 4} \right].
\end{eqnarray}

\section{Revision of Moon's work}
\label{sec:revision_Moon}

Here we revise Moon's pioneering work using the derivations above. We first interpret Moon's formula and try to identify Moon's calculations for the values of $\sexpetw$ (table \ref{table:types_probs_sphere}). Then, we compare Moon's results with ours in order to obtain a formalization of the deviation between Moon's formulae from the actual value of the variance (sections \ref{sec:revision_Moon:complete} and \ref{sec:revision_Moon:complete_bipartite}).

Notice that (\ref{eq:var_C:rsa:complete:Moons_raw}) follows the pattern of (\ref{eq:var_C:general}) with some differences. An obvious one is that $q$ has been factored out. Therefore, it is convenient to rewrite the general formula of $\gvar{\cross}$ (\ref{eq:var_C:general}) equivalently as 
\begin{eqnarray*}
\gvar{\cross} = q \sum_{\omega\in\Omega} \frac{f_\omega}{q} \gexpetw.
\end{eqnarray*}
The values of $f_\omega/q$ for a complete graph are summarized in table \ref{table:freq_types_normalized} (see \ref{sec:relative_frequencies_of_types} for a detailed derivation), and allow one to hypothesize that (\ref{eq:var_C:rsa:complete:Moons_raw}) is actually showing
\begin{equation}
\label{eq:hypothesis}
\Msvar{\cross(\complete)}
	= q
	\left[
		\frac{f_{24}}{q} \sexpet{24} +
		\frac{f_{12}}{q} \sexpet{12} +
		\frac{f_{13}}{q} \sexpet{13} +
		\frac{f_{04}}{q} \sexpet{04}
	\right]
\end{equation}
with
\begin{eqnarray*}
\Msexpet{24} &= \frac{7}{64} \\
\Msexpet{04} &= -\frac{1}{64} \\
\Msexpet{12} &= \Msexpet{13} = \frac{\pi^2 - 8}{64 \pi^2}.
\end{eqnarray*}

Moon suggested that the product types $00$, $01$, $021$, $022$ and $03$ do not have any contribution in (\ref{eq:var_C:general}). On the one hand, the absence of types $00$ and $01$ could be explained by the fact that $\gexpet{00}=\gexpet{01}=0$ \cite{Alemany2018a}. However, as we have shown in section \ref{sec:var_C:rsa_ours}, the other types, i.e., $021$, $022$ and $03$, satisfy $\sexpetw \neq 0$. On the other hand, the value of $\Msexpet{13}$ differs from ours (table \ref{table:types_probs_sphere}). In the next two sections, we show  that Moon's values lead to inaccurate arithmetic expressions for $\svar{\cross(\complete)}$ and $\svar{\cross(\compbip)}$.

In order to validate our hypothesis (\ref{eq:hypothesis}), 
we can instantiate (\ref{eq:var_C:general:complete}) with the values $\Msexpetw$. Doing so, we obtain $\Msvar{\cross(\complete)}$ in (\ref{eq:var_C:rsa:complete:Moons_raw}). It is possible to repeat the same analysis for complete bipartite graphs (\ref{eq:var_C:rsa:complete_bipartite:Moons_raw}) but it is more complex because Moon gave it in this compact form (without the intermediate results he showed for complete graphs). However, we can still observe that $q$ was factored out in (\ref{eq:var_C:rsa:complete_bipartite:Moons_raw}).

\begin{table}
	\caption{Summary of the frequency of every product type in complete graphs. $f_\omega$, the frequency of product type $\omega$, and $f_\omega/q$, the ratio between its frequency and $q$, the size of the set of independent edge pairs. The values of the $f_\omega$ for complete and complete bipartite graphs can be found in table \ref{table:freqs:K__and__K_n1n2}.}
	\label{table:freq_types_normalized}
	\begin{indented}
	\item[]
	\begin{tabular}{lll}
		\br
		$\omega$ & $f_\omega(\complete)/|Q(\complete)|$ & $f_\omega(\compbip)/|Q(\compbip)|$ \\
		\mr
		00  & $3{n - 4 \choose 4}$ &  $2{n_1 - 2 \choose 2} {n_2 - 2 \choose 2}$ \\
		01  & $12{n - 4 \choose 3}$ & $4(n_1-2)(n_2 - 2)(n_1 + n_2 - 6)$ \\
		021 & $4(n-4)(n-5) = 8{n - 4 \choose 2}$ & $4(n_1-2)(n_2-2)$ \\
		022 & $4(n-4)(n-5) = 8{n - 4 \choose 2}$ & $(n_2 + n_1 -  5)(n_2 + n_1 - 4)$ \\
		03  & $8(n-4)$ &  $2(n_1 + n_2 - 4)$ \\
		04  & $2$ & $1$ \\
		12  & $(n-4)(n-5) = 2{n - 4 \choose 2}$ & $2(n_1 - 2)(n_2 - 2)$ \\
		13  & $4(n-4)$ & $2(n_1 + n_2 - 4)$ \\
		24  & $1$ & $1$ \\
		\br
	\end{tabular}
	\end{indented}
\end{table}

\subsection{The variance of $\cross$ in complete graphs}
\label{sec:revision_Moon:complete}

First, we approximate the deviation of $\Msvar{\cross(\complete)}$ (\ref{eq:var_C:rsa:complete:Moons_raw}) from the actual value of the variance (\ref{eq:var_C:rsa:complete:ours})
\begin{align}
\label{eq:revision_Moon:var_C_complete:deviation}
\svar{\cross(\complete)} - \Msvar{\cross(\complete)}
	\approx
	3{n \choose 4}(-7\cdot 10^{-5} (n - 4)(n - 122.48)) = \bigO{n^6}.
\end{align}

\begin{figure}
	\centering
	\includegraphics[scale=0.6]{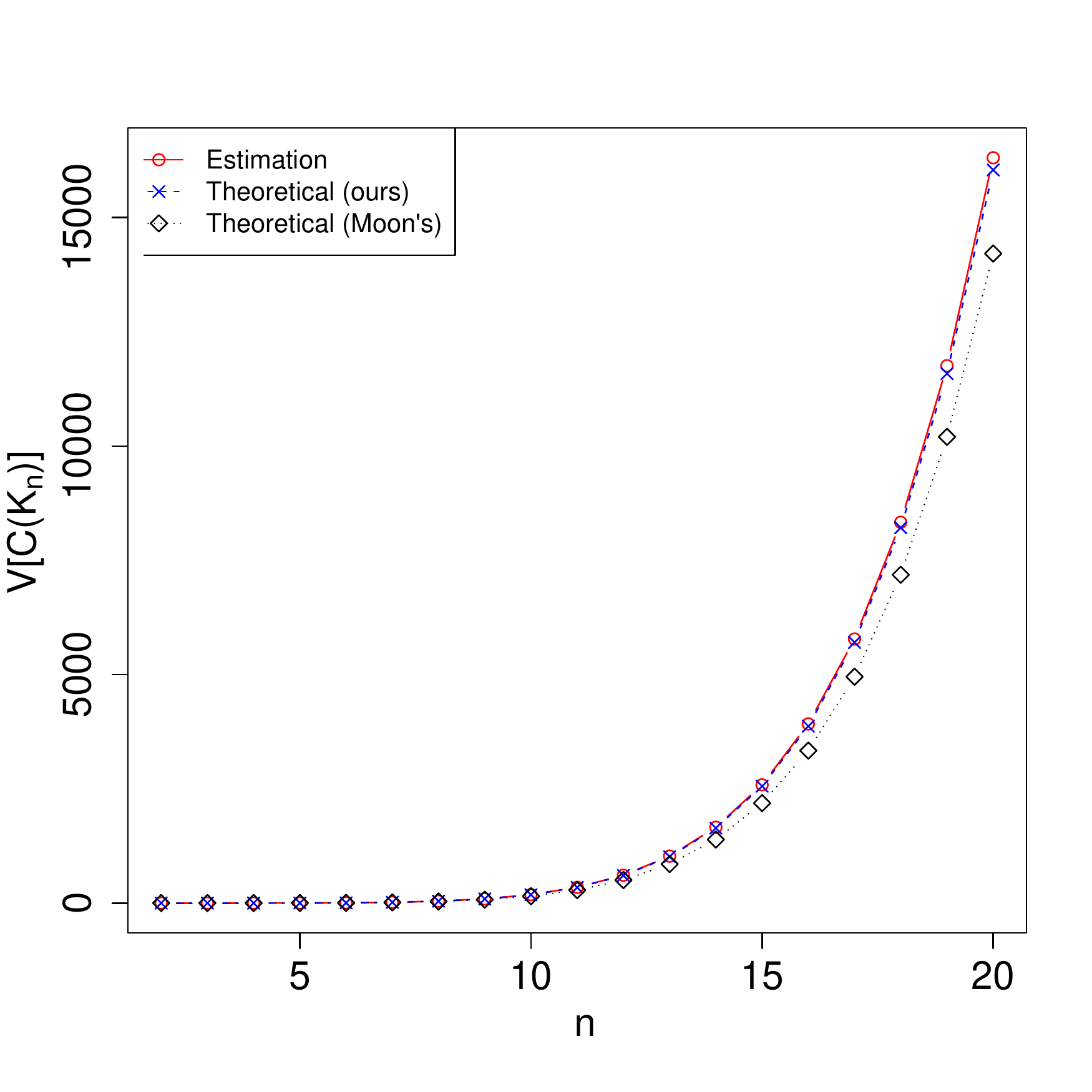}
	\caption{Comparison of $\Msvar{\cross(\complete)}$ (\ref{eq:var_C:rsa:complete:Moons_raw}, black diamonds), our proposal of $\svar{\cross(\complete)}$ (\ref{eq:var_C:rsa:complete:ours}, blue crosses), and $\asvar{\cross(\complete)}$, a numerical estimate obtained via computer simulations (red circles) using $N=10^7$ random arrangements for each $n$.}
	\label{fig:revision_Moon_complete_graphs}
\end{figure}

The accuracy of (\ref{eq:var_C:rsa:complete:ours}) in predicting $\svar{\cross(\complete)}$ is shown in figure \ref{fig:revision_Moon_complete_graphs}. Our prediction fits extremely well the numerical estimate of $\svar{\cross(\complete)}$ obtained via computer simulations (section \ref{sec:methods}) while $\Msvar{\cross(\complete)}$ deviates from $\svar{\cross(\complete)}$. Upon solving the equation $\svar{\cross(\complete)} - \Msvar{\cross(\complete)}=0$ for $n$, we can see that such deviation takes the form of underestimation for $n \lesssim 123$, and overestimation for $n\gtrsim 123$. This is why figure \ref{fig:revision_Moon_complete_graphs} shows only underestimation.

\subsection{The variance of $\cross$ in complete bipartite graphs}
\label{sec:revision_Moon:complete_bipartite}

First, we approximate the deviation of $\Msvar{\cross(\compbip)}$ (\ref{eq:var_C:rsa:complete_bipartite:Moons_raw}) from the actual value of the variance (\ref{eq:var_C:rsa:complete_bipartite:ours})
\begin{align}
\label{eq:revision_Moon:var_C_complete_bipartite:deviation}
\svar{\cross(\compbip)} &- \Msvar{\cross(\compbip)} \nonumber\\
&\approx
	\frac{1}{16}{n_1 \choose 2}{n_2 \choose 2}
	\left[
		3 - \frac{1}{\pi^2}((n_1 - 1)(n_2 - 1)(\pi^2 - 8) + 2(\pi^2 - 4))
	\right] \nonumber\\
&- 0.00068{n_1 \choose 3}{n_2 \choose 3} + 0.12\left[ {n_1 \choose 3}{n_2 \choose 2} + {n_1 \choose 2}{n_2 \choose 3} \right] \nonumber\\
&\phantom{=}  + 0.071\left[{n_1 \choose 4}{n_2 \choose 2} + {n_1 \choose 2}{n_2 \choose 4} \right] \nonumber\\
&= \bigO{n_1^2n_2^2(n_1^2 + n_2^2 + n_1 + n_2 - n_1n_2)}.
\end{align}

\begin{figure}
	\centering
	\includegraphics[scale=0.4]{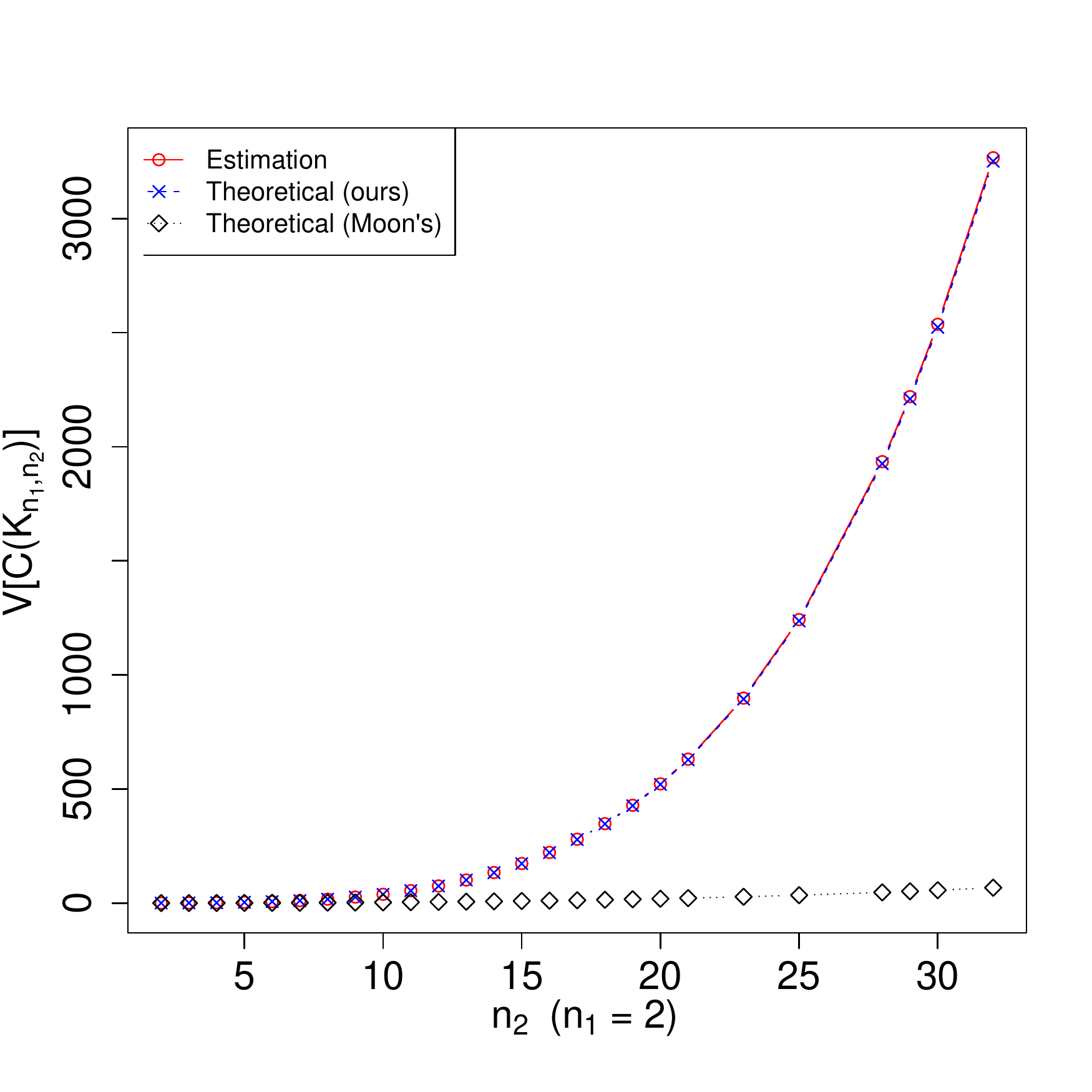}
	\includegraphics[scale=0.4]{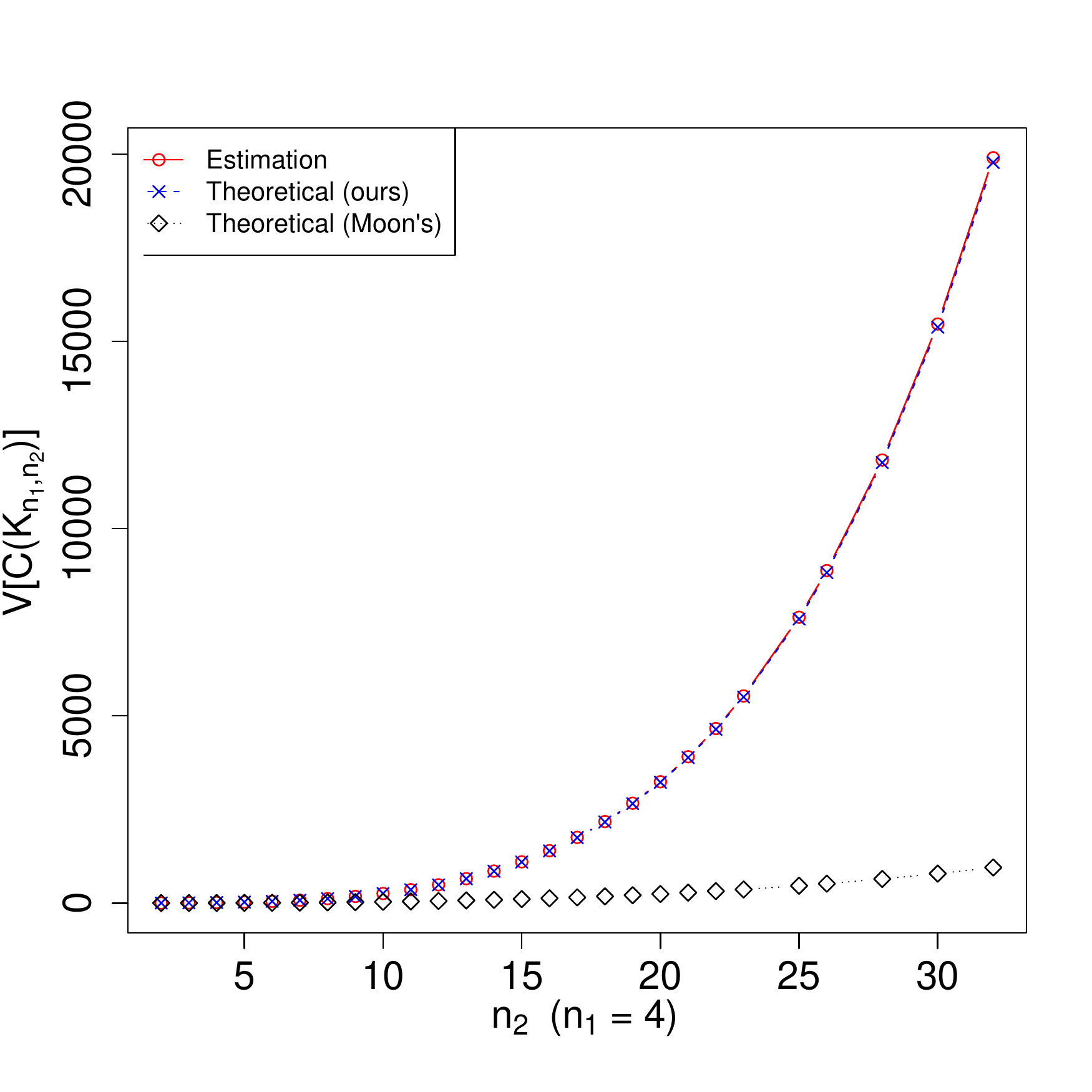}
	\\
	\includegraphics[scale=0.4]{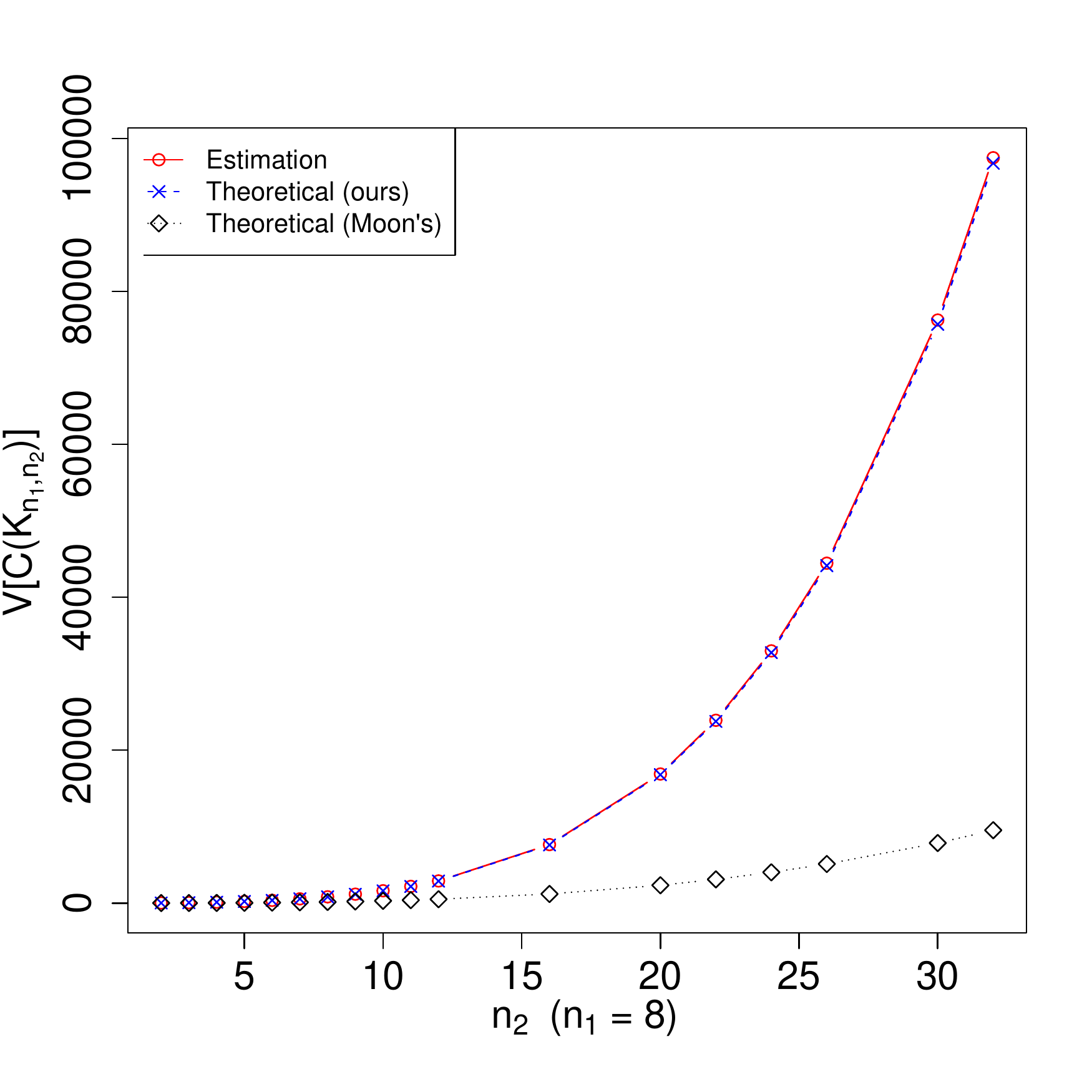}
	\includegraphics[scale=0.4]{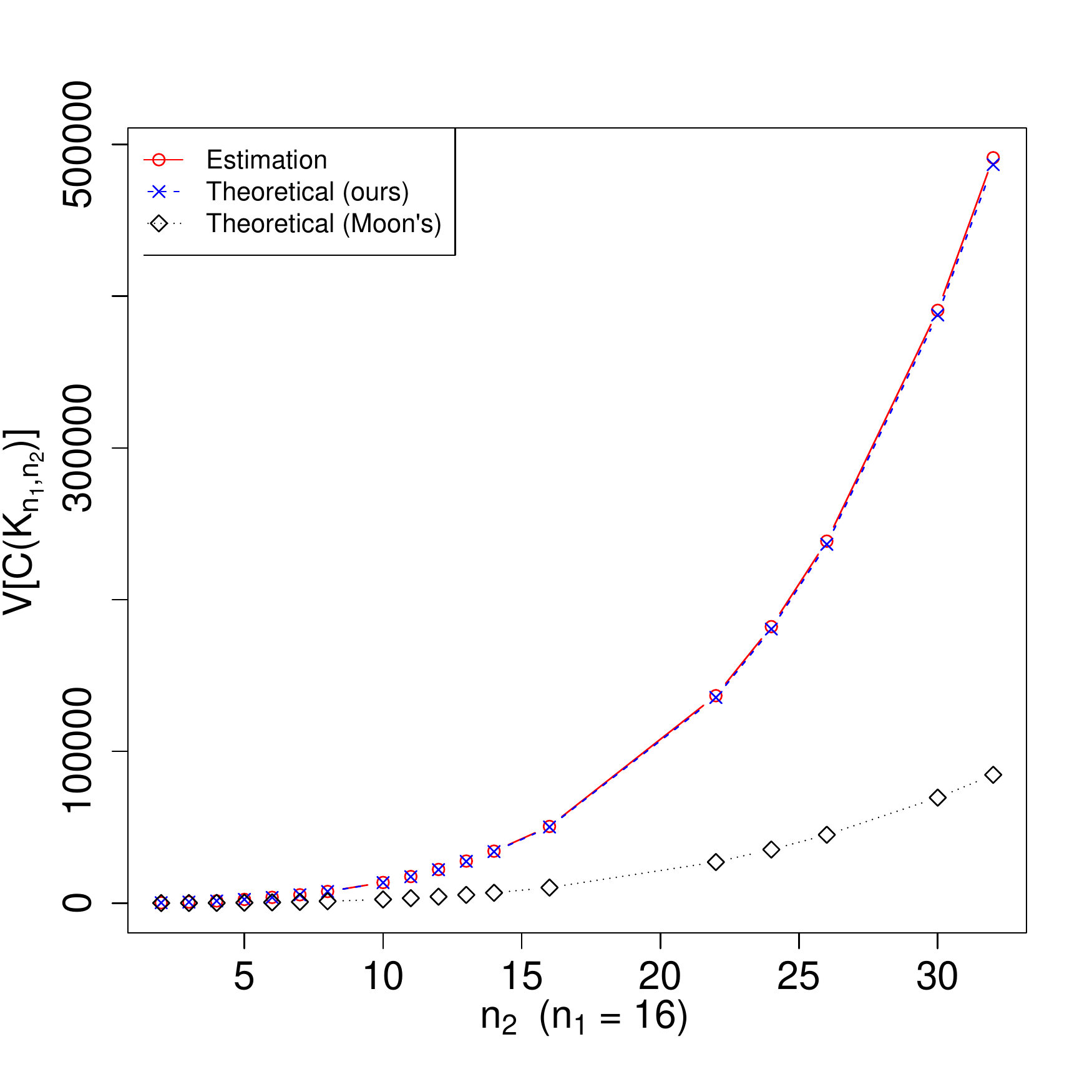}
	\caption{Comparison of $\Msvar{\cross(\compbip)}$ (\ref{eq:var_C:rsa:complete_bipartite:Moons_raw}), our proposal of $\svar{\cross(\compbip)}$ (\ref{eq:var_C:rsa:complete_bipartite:ours}), and $\asvar{\cross(\compbip)}$, a numerical estimate obtained via computer simulations using $N=10^7$ random arrangements for each pair $n_1,n_2$. $\svar{\cross(\compbip)}$ denotes the variance of the number of edge crossings in random spherical arrangements of a complete bipartite graph as a function of $n_1$, with $n_2$ fixed. $n_1$ and $n_2$ are, respectively, the number of vertices of the first and second partitions.}
	\label{fig:var_C:complete_bipartite:distribution}
\end{figure}

The accuracy of (\ref{eq:var_C:rsa:complete_bipartite:ours}) in predicting $\svar{\cross(\compbip)}$ is shown in figure \ref{fig:var_C:complete_bipartite:distribution}. Again, our prediction fits extremely well the numerical estimate of $\svar{\cross(\compbip)}$ (section \ref{sec:methods}) while $\Msvar{\cross(\compbip)}$ deviates from $\svar{\cross(\compbip)}$ as expected from (\ref{eq:revision_Moon:var_C_complete_bipartite:deviation}).

\begin{figure}
	\centering
	\includegraphics[scale=0.3]{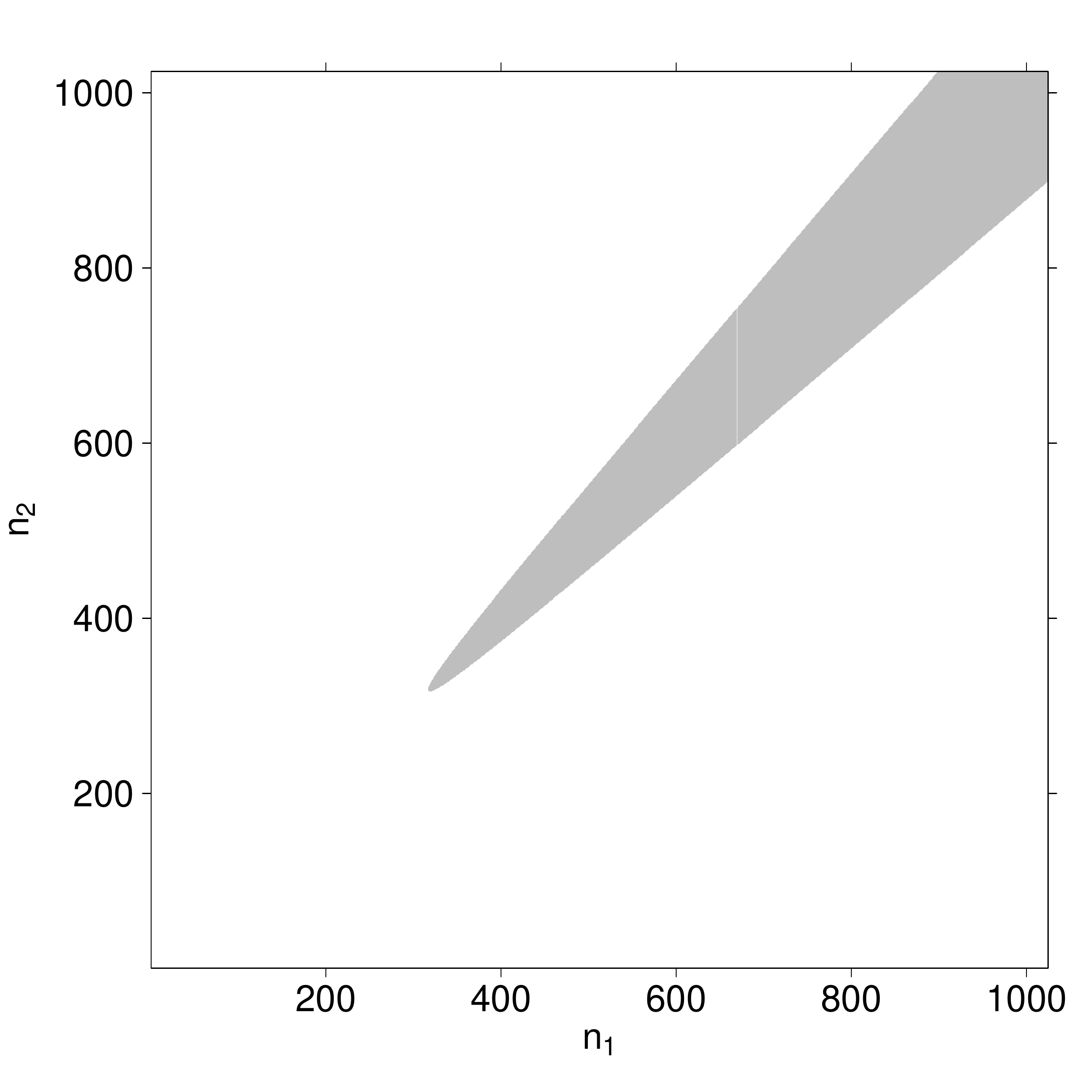}
	\caption{The sign of $\svar{\cross(\compbip)} - \Msvar{\cross(\compbip)}$ as a function of $n_1$ and $n_2$ (white for positive and gray for strictly negative). }
  \label{fig:sign_of_deviation_bipartite}
\end{figure}

The analysis of \ref{eq:revision_Moon:var_C_complete_bipartite:deviation} shows that Moon's equation underestimates the true variance unless $n_1$ and $n_2$ are sufficiently large and not too far from each other; overestimation occurs otherwise (figure \ref{fig:sign_of_deviation_bipartite}). Accordingly, figure \ref{fig:var_C:complete_bipartite:distribution} shows only understimation.

\section{Discussion}
\label{sec:discussion}

Back in 1965, Moon studied the variance of edge crossings in spherical arrangements of graphs, $\svar{\cross}$. In this article, we have revised Moon's work in light of recent discoveries on $\gvar{\cross}$, the variance in general layouts \cite{Alemany2018a}. We have applied them to derive an arithmetic expression for $\svar{\cross}$, which consisted of calculating the values for $\sexpetw$, the expectation of the types of products summarized in table \ref{table:types_probs_sphere}, and then deriving expressions for the variance in complete graphs (\ref{eq:var_C:rsa:complete:ours}) and in complete bipartite graphs (\ref{eq:var_C:rsa:complete_bipartite:ours}).

While some of the $\sexpetw$ we have calculated are in agreement with Moon's results, i.e. $\sexpetw$ for $\omega\in\{00,01,04,12,24\}$, we have found that others are not, i.e. $\omega\in\{021,022,03,13\}$. Our belief that the calculation of $\sexpet{13}$ by Moon is inaccurate is supported by the analyses in section \ref{sec:revision_Moon} and the values that were obtained in section \ref{sec:var_C:rsa_ours} (summarized in table \ref{table:types_probs_sphere}). Moreover, it appears that Moon considered some of these type's contribution to the variance to be null, i.e., $\sexpetw=0$ for $\omega\in\{021,022,03\}$, supported by the claim that \cite{Moon1965a} ``{\em the two variables appearing are independent and hence the expectation of their product equals the product of their individual expectations, i.e.,  zero}.'' We have found both via computer simulations and analytical calculations that this is not the case, hence proving that Moon's formula for $\svar{\cross}$ is inaccurate, as shown in sections \ref{sec:revision_Moon:complete} and \ref{sec:revision_Moon:complete_bipartite}.

We are aware of an erratum of Moon's article \cite{Moon1977a} where it is acknowledged that the initial derivation might be inaccurate and provides two arithmetic expressions for $\svar{\cross}$, i.e.,
\begin{eqnarray}
\svar{\cross(\complete)}
	&= \frac{\pi^2 - 8}{512\pi^2}n^6 + \bigO{n^5} \label{Moon1977_unipartite_equation} \\
\svar{\cross(\compbip)}
	&= \frac{\pi^2 - 8}{128\pi^2}n_1^2n_2^2(n_1^2 + n_2^2) + \bigO{n_1^2n_2^2(n_1 + n_2)}. \label{Moon1977_bipartite_equation}
\end{eqnarray}
Although \cite{Moon1977a} does not provide any explanation concerning why the initial derivation was inaccurate, our asymptotic equations (\ref{eq:revision_Moon:var_C_complete:deviation})-(\ref{eq:revision_Moon:var_C_complete_bipartite:deviation}) are consistent with Moon's (\ref{Moon1977_unipartite_equation})-(\ref{Moon1977_bipartite_equation}).

In this work, besides having contributed with an accurate expression for $\svar{\cross}$, we have also shed light on the origins of the bias in Moon's formulae. Moreover, the values of $\sexpetw$ coupled with the arithmetic expressions of the $f_\omega$ \cite{Alemany2018a} pave the way towards the obtention of $\svar{\cross}$ in other types of graphs.

Our current revision of \cite{Moon1965a} offers various possibilities for future research. From the exact calculation of $\sexpe{\gamma_\omega}$ for $\omega \in \{021, 022, 03, 13\}$, to the main goal of \cite{Moon1965a}, that was to show that the distribution of $\cross$ is asymptotically normal.

It is worth noting that the computational resources needed to test the mathematical arguments and results of \cite{Moon1965a} were not available at that time. Crucially, they are needed to approximate numerically the values of $\sexpe{\gamma_\omega}$ for $\omega\in\{021,022,03,13\}$ (section \ref{sec:methods}). Perhaps not so surprisingly, computers are helping to validate and improve classic work from the 1960-70s (e.g., \cite{Esteban2015a}). We hope that our research stimulates further research on crossings in spherical arrangements and other layouts.

\section{Methods}
\label{sec:methods}

Here we explain a way to generate points uniformly at random (u.a.r.) on the surface of the sphere (section \ref{sec:methods:random_points}), to determine if two arcs intersect (section \ref{sec:methods:arc_arc_test}), to speed up the intersection test (section \ref{sec:methods:arc_arc_test:speedup}), to estimate the values of $\sprobalphastw$ and $\sexpetw$ (section \ref{sec:methods:estimate_exp_types}), and, finally, a way to estimate $\svar{\cross}$ (section \ref{sec:methods:estimate_var_in_graph}).

\subsection{Generating random points}
\label{sec:methods:random_points}

Generating points uniformly at random on the surface of a unit sphere was done by generating random polar coordinates following \cite{Weisstein2003a}. First, we generated random values for variables $u,v\in [0,1]$. Then we generated the polar coordinates with
\begin{equation*}
\theta = 2\pi u, \quad
\phi = \cos^{-1} (2v - 1),
\end{equation*}
and, finally, the coordinates $(x,y,z)$ of a point were calculated using
\begin{equation*}
x = \sin{(\theta)}\cos{(\phi)}, \quad
y = \sin{(\theta)}\sin{(\phi)}, \quad
z = \cos{(\theta)}.
\end{equation*}

The random values for generating points uniformly at random on the surface of a sphere were generated using the C++'s header \textit{random}: we used the \textit{default\_random\_engine} initialized with the C++'s \textit{random\_device}, a non-deterministic random number generator that uses hardware entropy source. This is used to seed the random engine once for each set of tests, that is, for a whole set of replicas. The \textit{default\_random\_engine} is then used in the \textit{uniform\_real\_distribution} object to generate floating-point pseudo-random numbers uniformly at random.

\subsection{Arc-arc intersection test}
\label{sec:methods:arc_arc_test}

\begin{figure}
	\centering
	\includegraphics[scale=0.5]{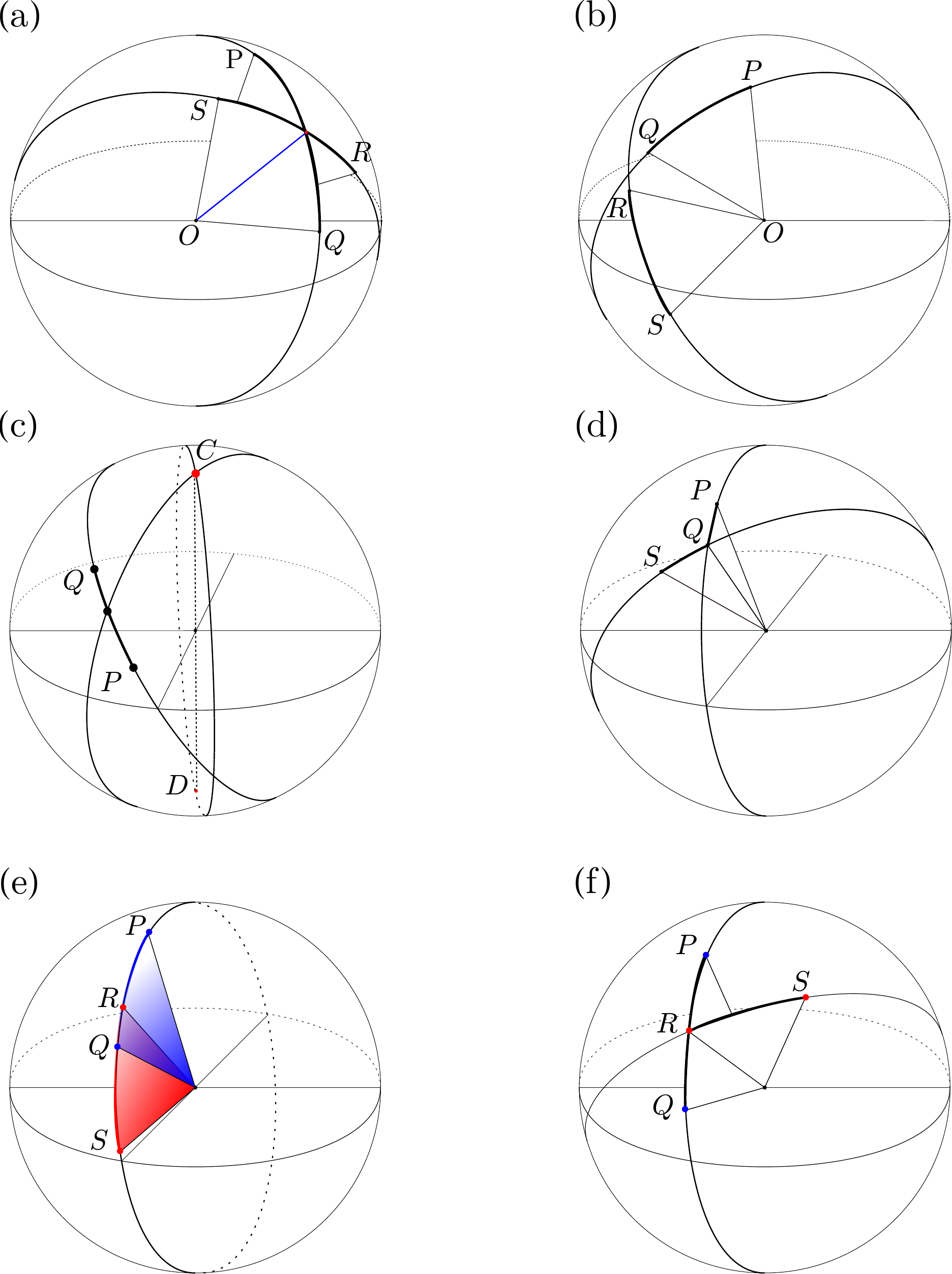}
	\caption{The different scenarios in which two great arcs might arcs cross. (a) Two arcs that cross. Their triangles intersect at a segment. (b) Two arcs that cross. Their triangles intersect only at point $O$. (c) Diametrically opposed points ($\cross$ and $D$, in red), two possible great circles through $\cross$ and $D$. One of the great circles crosses great arc $PQ$. (d) Two arcs sharing an endpoint. (e) 4 points lying on the same great circle (overlapping): intersection between the triangle of the arc $PQ$ (blue) and the arc $RS$ (red) at an infinite number of points. (f) 3 points lying on the same great circle. Intersection at a single point.}
	\label{fig:all_crossings}
\end{figure}

In \cite{Moon1965a}, Moon mentioned in passing possible ambiguities in the definition of edge crossing ``{\em arising when different vertices coincide, when two vertices are diametrically opposed, when more than two vertices lie on the same great circle, or when more than two arcs intersect at the same point may be disregarded as they occur only with probability zero}.'' The situation that ``{\em more than two arcs intersect at the same point}'', as explained in the introduction when introducing the generic arrangement $*$, contributes to $\cross$ with ${e \choose 2}$ crossings, where $e$ is the number of arcs. The other two cases, still have probability zero but can appear when calculating arc-arc intersections computationally due to lack of numerical precision. Below we cover them in a formalization of the notion of arc-arc intersection. 

Let $a_1=(S,T)$ and $a_2=(U,V)$ be the two arcs with points $S,T,U,V$ pairwise distinct. Figure \ref{fig:all_crossings}(a) illustrates an example of two arcs that cross and figure \ref{fig:all_crossings}(b) two that do not cross. We adopt the following operational definition of arc intersection: arcs $a_1$ and $a_2$ intersect if, and only if, the triangles $t_1=(O,S,T)$, $t_2=(O,U,V)$ intersect at some point other than $O$.

In \ref{fig:all_crossings}(a) the triangles intersect at a segment while in figure \ref{fig:all_crossings}(b), the triangles intersect in only at point $O$. If the intersection between the triangles is a single point then it must be $O$. There is an extreme case of intersection only at the origin that is worth mentioning: when an arc has its two endpoints diametrically opposed to each other (figure \ref{fig:all_crossings}(c)) as pointed out by Moon. That is, the shortest Euclidean distance between them is exactly the diameter of the sphere. Such arc defines a degenerate triangle (actually a segment) that intersects the triangle of the other arc at $O$. {\em A priori}, any second arc, with its endpoints diametrically opposed or not, always intersects the first. However, we do not consider this as a valid intersection due to our operational definition. In our implementation of the test, this situation is detected by testing whether any of the triangles $t_1$ or $t_2$ is degenerated, i.e., by testing whether $t_1$ or $t_2$ is actually a line segment. 

Now we review two other situations considered by Moon that have probability zero in relation to our framework:
\begin{itemize}
	\item The situation that {\em ``more than two vertices lie on the same great circle''} can happen in two circumstances: when four points lie on the same great circle (the intersection between the two triangles is another triangle) then the arcs may overlap (figure \ref{fig:all_crossings}(e)) and also when three points lie on the same great circle one of the endpoints of one arc may be between the endpoints of the other arc (figure \ref{fig:all_crossings}(f)).

	\item When the intersection between the triangles is a segment then the arcs intersect at exactly one point, which might be one of the endpoints. Figure \ref{fig:all_crossings}(a) and figure \ref{fig:all_crossings}(f) shows two arcs crossing, not at an endpoint. Figure \ref{fig:all_crossings}(d) illustrates {\em ``when different vertices coincide''}, namely an arc-arc intersection at an endpoint.
\end{itemize}

We implemented the arc-arc intersection test using the CGAL library \cite{CGAL_library} (version 4.11.1). Computations were done using CGAL's kernel `Exact\_predicates\_exact\_constructions\_kernel'.

\subsection{Arc-arc intersection test speed-up}
\label{sec:methods:arc_arc_test:speedup}

The kernel above provides a highly precise but time-consuming arc-arc intersection test. Because of the latter, we run a fast test to filter out those pairs of arcs that do not cross. This test is based on three sufficient conditions for non-intersection. If these conditions fail, then we run the time-consuming intersection test provided by CGAL. 

The sufficient conditions are based on a division of the sphere's surface into $8$ octants. Each octant is defined by the sign of the coordinates of the points in them. The octant of a point $A=(x, y, z)$ is $\oct(A) = (\sgn(x), \sgn(y), \sgn(z))$, where $\sgn$ is the sign function, i.e. 
\begin{equation*}
\sgn(x) = \left\{ 
         \begin{array}{c}
            \frac{x}{|x|} \mbox{~if~} x \neq 0 \\
            1 \mbox{~if~} x = 0.
         \end{array} 
         \right.
\end{equation*}

We now define the three sufficient conditions for non-intersection. Any pair of disjoint arcs $ST$ and $UV$ does not cross
\begin{itemize}
	\item If the arcs fall into different octants, namely $\oct(S) = \oct(T) \neq \oct(U) = \oct(V)$, then the arcs cannot intersect.

	\item If $ST$ and $UV$ are separated by one of the planes $x=0$, $y=0$, or $z=0$. Let $\oct_c(P)=\sgn(c)$ for $c\in\{x,y,z\}$ of a point $P=(x,y,z)$. Formally, two arcs $ST$ and $UV$ are separated by one of the planes $x=0$, $y=0$, $z=0$ when there exists a coordinate $c$ such that $\oct_c(S) = \oct_c(T) \neq \oct_c(U) = \oct_c(V)$.

	\item If points $U,V$ are located in the same half space defined by the plane through points $S,T,O$ then the arcs do not intersect.
\end{itemize}

In order to achieve a higher reduction of the amount of arc-arc intersection tests, we apply the first two sufficient conditions defined above to $3l$, $l\ge 1$, transformations of the original set of points $\mathcal{S}$. The $i$th group of three transformations consists of rotations of all points around each axis separately by angle $\theta_i$. More precisely, the set of $3l$ transformations of $\mathcal{S}$ is
\begin{equation*}
\{ R_{\theta_i}^{OX}(\mathcal{S}), R_{\theta_i}^{OY}(\mathcal{S}), R_{\theta_i}^{OZ}(\mathcal{S}) \}_{i=1}^l
\end{equation*}
where $R_{\theta_i}^e(\mathcal{S}) = \{R_{\theta_i}^e(P) \;|\; P\in \mathcal{S}\}$ is the rotation of all points in $S$ around axis $e$ by an angle $\theta_i$. The angles $\theta_i$ are defined as $\theta_i = i\pi/2(l+1)$ for all $i \in \{1,...,l\}$. If one of the sufficient conditions defined above is true for a given pair of independent arcs in some of the transformations then the arcs do not intersect.

\subsection{Estimating the expectation of types}
\label{sec:methods:estimate_exp_types}

The integrals used to calculate the values of the $\sprobalphastw$ (section \ref{sec:var_C:rsa_ours}) have been approximated numerically using the computer algebra system Maple\textsuperscript{TM}\footnote{Maple is a trademark of Waterloo Maple Inc.} \cite{Maple2018}. We used the function \textit{integrate} with the parameter \textit{method} set to \textit{CubaDivonne} and \textit{CubaCuhre} depending on the case.

Each $\sprobalphastw$ was estimated numerically by generating $T=10^7$ random spherical arrangements (section \ref{sec:methods:random_points}) of a $\complete[10]$. This implies that each $\sprobalphastw$ was estimated over $T f_\omega$ replicas, where $f_\omega$ is given in table \ref{table:freqs:K__and__K_n1n2}.

For each random layout, we calculated what pairs of arcs intersected (sections \ref{sec:methods:arc_arc_test} and \ref{sec:methods:arc_arc_test:speedup}). Then we classified each pair of independent arcs, i.e., each of the elements in $Q \times Q$, and used this information to compute the $\sprobalphastw$ and then $\sexpetw$.

Our simulations confirm the correctness of the results obtained in section \ref{sec:var_C:rsa_ours}. Our $T$ simulations yielded the results presented in the middle columns of table \ref{table:results_simulation_exp_gammas}. Since the values were obtained by classifying the different elements of $Q\times Q$ in a complete graph, each type was sampled at different amounts. This means that the precision at which they were obtained is different for every type, which we convey in the rightmost column of table \ref{table:results_simulation_exp_gammas}. That column indicates the amount of iterations for which the last decimal of the estimate of $\sprobalphastw$ did not change before reaching the end of the simulation.

\begin{table}
	\caption{Numerical estimates of $\sprobalphastw$ and $\sexpetw$ obtained via computer simulations, for all types $\omega\in\Omega$ (\ref{eq:alpha_products:code_types}). Values $\asprobalphastw$ have been truncated to the last decimal that did not change in the last amount iterations indicated in the column ``Iterations'' (a ``-'' indicates that the value remained the same throughout the whole simulation). $f_\omega$ is the amount of products of type $\omega$ that can be found in $\complete[10]$ (they are obtained from table \ref{table:freqs:K__and__K_n1n2} with $n=10$).  The number of samples used to calculate the numerical estimates is $Tf_\omega$. }
	\label{table:results_simulation_exp_gammas}
	\begin{indented}
	\item[]
	\begin{tabular}{lllll}
	\br
	$\omega$	& $\asprobalphastw$	& $\asexpetw$	& $f_\omega$	& Iterations	\\
	\mr
	00			& 0.0156253					&  0.0000003				& 28350			& 4750		\\
	01			& 0.0156258					&  0.0000008				& 151200		& 5750		\\
	021			& 0.0126703					& -0.0029546				& 75600			& 5600		\\
	022			& 0.0185812					&  0.0029562				& 75600			& 5550		\\
	03			& 0.010417					& -0.005207					& 30240			& 49900		\\
	04			& 0.0000000					& -0.0156250				& 1260			& -			\\
	12			& 0.01858					&  0.00295					& 18900			& $>$40000	\\
	13			& 0.0312507					&  0.0156257				& 15120			& 3650		\\
	24			& 0.125001					&  0.109376					& 630			& 39250		\\
	\br
	\end{tabular}
	\end{indented}
\end{table}

\subsection{Estimating $\svar{\cross}$ in a graph}
\label{sec:methods:estimate_var_in_graph}

Estimating $\svar{\cross}$ on a graph consists of
\begin{enumerate}
	\item Generating $N$ random spherical layouts (section \ref{sec:methods:random_points}),

	\item Calculating the value of $\cross$ for each layout (using the arc-arc intersection test described in section \ref{sec:methods:arc_arc_test} , optionally with the improvements explained in section \ref{sec:methods:arc_arc_test:speedup}),
	
	\item Applying an unbiased estimator of variance to the $N$ values of $\cross$. 
\end{enumerate}

However, for large values of $n$ (in complete graphs), or large values of $n_1$ and $n_2$ (in complete bipartite graphs), estimating the variance with $N=10^7$ replicas turned out to be a rather time-consuming task. Therefore, for large $n$ (or $n_1$ and $n_2$), the first two steps were organized into partitions and then parallelized (each partition was in charge of the processing a certain number of random layouts).

\section*{Acknowledgments}

We are greatly indebted to J. W. Moon for his comments on this manuscript and further clarifications. We are grateful to Kosmas Palios for making us aware of Moon's work and for helpful discussions. We also thank Vera Sacrist{\'a}n for valuable advice. RFC and LAP were supported by the grant TIN2017-89244-R from MINECO (Ministerio de Econom{\'i}a y Competitividad), the acknowledgement 2017SGR-856 (MACDA) from AGAUR (Generalitat de Catalunya), and MM was supported by MTM2015-63791-R (MINECO/FEDER), Gen. Cat. DGR 2017SGR1336 and H2020-MSCA-RISE project 734922-CONNECT.

\appendix
\addtocontents{toc}{\fixappendix}
\setcounter{section}{0}

\section{Relative frequencies of types}
\label{sec:relative_frequencies_of_types}

We recall the definition of the falling factorial, i.e. \cite{Bollobas1998a}
\begin{eqnarray*}
(n)_{x}
	&= n(n-1)(n-2)...(n-x+1) = x! {n \choose x}.
\end{eqnarray*}
We derive $f_\omega/q$ for all types of products with the help of the values of $f_\omega$ for complete graphs and complete bipartite graphs (table \ref{table:freqs:K__and__K_n1n2}) and a property of the quotient of binomial coefficients, namely
\begin{equation*}
\frac{{n \choose x}}{{n \choose y}} = (n-y)_{x-y} \frac{y!}{x!}. 
\end{equation*}
when $x \geq y$. The results are summarized in table \ref{table:freq_types_normalized}.

\subsection{Complete graphs}

First,
\begin{eqnarray*}
\frac{f_{00}}{q}
	&= \frac{630{n \choose 8}}{3{n \choose 4}}
	 = \frac{1}{8} (n-4)_4
	 = \frac{4!}{8} {n-4 \choose 4}
	 = 3 {n-4 \choose 4}. 
\end{eqnarray*}
Second, 
\begin{eqnarray*}
\frac{f_{01}}{q}
	&= \frac{1260{n \choose 7}}{3{n \choose 4}}
	 = 2 (n-4)_3
	 = 2 \cdot 3! {n-4 \choose 3}
	 = 12 {n-4 \choose 3}
\end{eqnarray*}
\begin{eqnarray*}
\frac{f_{021}}{q} = \frac{f_{022}}{q}
	&= \frac{360{n \choose 6}}{3{n \choose 4}}
     = 4(n-4)(n-5)
     = 8{n-4 \choose 2}. 
\end{eqnarray*}
As $f_{12} = f_{021}/4$, we also get
\begin{eqnarray*}
\frac{f_{12}}{q} &= (n-4)(n-5)
                  = 2{n-4 \choose 2}. 
\end{eqnarray*}
Fourth, 
\begin{eqnarray*}
\frac{f_{03}}{q} &= \frac{120{n \choose 5}}{3{n \choose 4}}
                  = 8(n-4).
\end{eqnarray*}
As $f_{13} = f_{03}/2$, we also get $f_{13}/q = 4(n-4)$.  Fifth, $f_{04}/q = 2$ and $f_{24}/q = 1$ trivially.  

\subsection{Complete bipartite graphs}

First,
\begin{eqnarray*}
\frac{f_{00}}{q}
	&= \frac{1}{2} (n-2)_{2} (m-2)_{2}
     = 2 {n - 2 \choose 2} {m - 2 \choose 2}.
\end{eqnarray*}
Second, the fact that 
\begin{eqnarray*}
\frac{{n \choose 4}{m \choose 3}}{q} &= \frac{1}{36}(n-2)_2 (m-2)
\end{eqnarray*}
gives
\begin{eqnarray*}
\frac{f_{01}}{q}
	&= 144 \left[ \frac{{n \choose 4}{m \choose 3}}{q} + \frac{{n \choose 3}{m \choose 4}}{q} \right]
	 = 4(n-2)(m-2) (n+m-6).
\end{eqnarray*}
Third, the fact that 
\begin{eqnarray*}
\frac{{n \choose 3}{m \choose 3}}{q} &= \frac{1}{18}(n-2)(m-2)
\end{eqnarray*}
gives
\begin{eqnarray*}
\frac{f_{021}}{q}
	&= 72 \frac{{n \choose 3}{m \choose 3}}{q}
     = 4(n-2)(m-2) 
\end{eqnarray*}
and 
\begin{eqnarray*}
\frac{f_{12}}{q}
	&= 36 \frac{{n \choose 3}{m \choose 3}}{q}
     = 2(n-2)(m-2). 
\end{eqnarray*}
Fourth, the fact that
\begin{eqnarray*}
\frac{{n \choose 2}{m \choose 4}}{q} &= \frac{1}{24}(m-2)_2
\end{eqnarray*}
and that $f_{12}$ is included in the definition of $f_{022}$ as third summand, gives 
\begin{eqnarray*}
\frac{f_{022}}{q}
	&= 24 \left[ \frac{{n \choose 2}{m \choose 4}}{q} + \frac{{n \choose 4}{m \choose 2}}{q} \right] + f_{022} \\
	&= (m-2)_2 + (n-2)_2 + 2(n-2)(m-2) \\
	&= \left\{
	\begin{array}{l}
	2\left[{m - 2 \choose 2} +{n - 2 \choose 2} + (n-2)(m-2)\right] \\
    (m + n - 5)(m + n - 4).
	\end{array}
    \right.
\end{eqnarray*}
Finally, one has  
\begin{eqnarray*}
\frac{f_{04}}{q} &= \frac{f_{24}}{q} = 1
\end{eqnarray*}
and also
\begin{eqnarray*}
\frac{f_{03}}{q} &= \frac{f_{13}}{q} = 2(n+m-4)
\end{eqnarray*}
trivially.

\section*{References}
\bibliographystyle{plain}

\newcommand{\beeksort}[1]{}

\end{document}